\documentclass[reprint, amsmath,amssymb,superscriptaddress, titlepage]{revtex4-2}
\usepackage{my_package}
\usepackage{enumerate}

\newcommand{\as}{\vec{u}^s}
\newcommand{\vs}{\vec{v}^s}
\newcommand{\vf}{\vec{v}^f}

\newcommand{\asx}{\vec{u}^s_{\parallel}}
\newcommand{\vsx}{\vec{v}^s_{\parallel}}
\newcommand{\vfx}{\vec{v}^f_{\parallel}}

\newcommand{\asr}{u^s_{r}}

\newcommand{\asz}{u_z^s}
\newcommand{\vsz}{v^s_z}
\newcommand{\vfz}{v^f_z}

\newcommand{\xx}{\vec{x}_{\parallel}}

\newcommand{\phif}{\phi^f}
\newcommand{\phis}{\phi^s}

\newcommand{\q}{q_{\mathrm{e}}}

\newcommand{\sxx}{\tens{\sigma}_{\parallel}}
\newcommand{\sxz}{\vec{\sigma}_{\perp}}
\newcommand{\szz}{\sigma_{zz}}

\newcommand{\Ixx}{\tens{I}_{\parallel}}
\newcommand{\Txx}{\tens{T}_{\parallel}}

\newcommand{\TT}{\vec{\mathcal{T}}}
\newcommand{\TTx}{\vec{\mathcal{T}}_{\parallel}}
\newcommand{\TTz}{\mathcal{T}_{z}}

\newcommand{\rcx}{\vec{r}^c_{\parallel}}

\newcommand{\ez}{\vec{e}_z}

\newcommand{\nx}{\nabla_{\parallel}}

\begin{document}

\title{Drying-induced stresses in poroelastic drops on rigid substrates}

\author{Matthew~G.~Hennessy}
\email{matthew.hennessy@bristol.ac.uk}
\affiliation{Department of Engineering Mathematics, University of Bristol, Ada Lovelace Building, Bristol, BS8 1TW, United Kingdom}
\author{Richard~V.~Craster}
\affiliation{Department of Mathematics, Imperial College London, South Kensington Campus, London, SW7 2AZ, United Kingdom}
\author{Omar~K.~Matar}
\affiliation{Department of Chemical Engineering, Imperial College London, South Kensington Campus, London, SW7 2AZ, United Kingdom}


\begin{abstract}
  We develop a theory for drying-induced stresses in sessile, poroelastic drops undergoing evaporation on rigid surfaces.  Using a lubrication-like approximation, the governing equations of three-dimensional nonlinear poroelasticity are reduced to a single thin-film equation for the drop thickness.  We find that thin drops experience compressive elastic stresses but the total in-plane stresses are tensile.  The mechanical response of the drop is dictated by the initial profile of the solid skeleton, which controls the in-plane deformation, the dominant components of elastic stress, and sets a limit on the depth of delamination that can potentially occur.  Our theory suggests that the alignment of desiccation fractures in colloidal drops is selected by the shape of the drop at the point of gelation.  We propose that the emergence of three distinct fracture patterns in dried blood drops is a consequence of a non-monotonic drop profile at gelation.  We also show that depletion fronts, which separate wet and dry solid, can invade the drop from the contact line and localise the generation of mechanical stress during drying. Finally, the finite element method is used to explore the stress profiles in drops with large contact angles.
\end{abstract}

\maketitle

\section{Introduction}
The drying behaviour of thin films and drops is important to a multitude of industries and applications. 
The presence of a particulate phase introduces  appreciable changes to the evaporative process and leads to hydrodynamic and mechanical instabilities, sometimes resulting in cracking.
While traditionally viewed as detrimental, the onset of cracking
can play
an advantageous role in a number of modern applications, from affordable medical diagnostics \cite{zang2019} to high-resolution, high-throughput nano-patterning \cite{kim2016}. 
With such a broad range of applications, there is a growing need for efficient and accurate modelling capabilities of the stresses accompanying the evaporation of particle-laden films and drops.
Yet considerable complexity arises in these drying systems from the delicate interplay between capillarity, thermocapillarity, heat and mass transfer, contact lines (undergoing pinning and de-pinning), and,
crucially, the formation of a poroelastic network, which controls the formation of cracks and their morphology.
In this paper, we focus on the development of a theory for drying-induced stresses in sessile, poroelastic drops undergoing evaporation on rigid surfaces.

Many important patterning behaviours manifest during the evaporation of droplets \cite{adachi1995} due to the interaction between evaporation-driven flows and the contact line~\cite{larson2014}. The commonly encountered `coffee-ring' stain is one such example, where evaporation in the presence of a non-volatile solute promotes the appearance of a distinctly inhomogeneous deposit. 
Deegan \etal\cite{deegan1997,deegan2000a,deegan2000b} first explained the origin of this effect by the presence of an increased evaporative flux at the contact line, coupled with a resultant capillary-induced restoration flow.

Crack formation in drying colloidal drops is thought to be a multi-step process
originating from the coffee-ring effect~\cite{chen2016, sobac2014}.
The radially outwards capillary flow transports colloids to the
pinned contact line where they accumulate due to weak, counteracting 
diffusive effects~\cite{moore2021}.  Gelation occurs once the particle concentration exceeds a critical value, resulting in the local transformation of the liquid drop into a poroelastic solid; the porous elastic material then has an elastic skeleton with interconnected pores containing fluid.
A gelation front consequently emerges from the contact line and
propagates towards the drop centre~\cite{sobac2011}.  
Mechanical stresses develop
in the gelled region because of the competing effects of evaporation-driven contraction of the solid skeleton and its adhesion to the substrate.  Cracks therefore emerge as a mechanism to relieve mechanical stress.

Experiments have shown that a myriad of fracture patterns can occur during droplet evaporation~\cite{giorgiutti2018}.  When an aqueous drop with silica nanoparticles dries, fractures typically nucleate at the contact line and travel radially inwards, following the gelation front.  These fractures divide the deposit into an array of `petals' that simultaneously delaminate from the substrate, resulting in the entire solid `blooming' into a structure that resembles a lotus flower~\cite{parisse1996, lilin2020}.  In the case of dried blood drops, the solid deposit often has an orthoradial fracture at the contact line, an annular region with several radial fractures, and a central zone with smaller-scale fractures with no preferred orientation~\cite{brutin2011}.  Extensive experimental research has been carried out to elucidate the dependence of the fracture pattern on the contact angle~\cite{carle2013, yan2021}, drying rate~\cite{zeid2013, zeid2013b, giorgiutti2014}, substrate deformability~\cite{lama2021}, and particle hydrophobicity~\cite{anyfantakis2017} and concentration~\cite{brutin2013, bourrianne2021}.  However, a detailed theoretical description of stress generation in drying colloidal drops is lacking, with treatments relying on scaling analyses~\cite{choi2020} or one-dimensional models that do not capture the evolving and non-uniform thickness of the drop~\cite{giorgiutti2014, osman2020}.

Our work will deploy three-dimensional nonlinear poroelasticity to build a comprehensive model of drying-induced mechanical stresses in drops with a  pre-existing solid structure. Moreover, a lubrication-like approximation will be invoked to systematically reduce the governing equations.  
Poroelasticity theory is based on the premise that the solid phase is arranged into a porous and deformation structure referred to as the `solid skeleton'. 
Biot developed the theory of poroelasticity~\cite{biot1941,biot1956,biot1962} to account for the two-way coupling between the deformation of the solid skeleton and fluid flow within the pore space.
The theory formalised by Biot assumes the deformation of the solid skeleton is infinitesimal and thus describes the solid skeleton as a linearly elastic material. When the solid deformation is no longer infinitesimal, one must derive poroelastic models in the framework of nonlinear elasticity~\cite{coussy2004, macminn2016}.
The lubrication approximation has been used in tandem with the theory of poroelasticity~\cite{jensen1994, barry2001} to study thin films for various applications such as CO$_2$ sequestration~\cite{hewitt2015}, imbibition~\cite{kvick2017}, soft contact~\cite{skotheim2004, skotheim2005}, and biomechanics~\cite{argatov2011,argatov2015}.

By combining poroelasticity with the lubrication approximation, we are able to provide novel physical insights into the internal droplet dynamics.  In the case of axisymmetric drops with circular contact lines, we find that the initial profile of the poroelastic drop plays a critical role in selecting the modes of in-plane deformation, thereby determining whether the radial or hoop stresses dominate.  Our work suggests that the fracture patterns appearing in drying colloidal drops are dictated by the shape of the drop at the point of gelation, in agreement with the experimental observations of Bourrianne \etal\cite{bourrianne2021}.  By comparing the magnitude of the radial and hoop stresses, we correctly predict the emergence of three distinct fracture patterns in dried drops of blood~\cite{sobac2011}.  We also find that the drop profile affects the depth of delamination, which may not reach the drop centre,
in line with the experiments of Osman \etal\cite{osman2020}.  We show that a sharp decrease in the permeability during drying can result in the formation of depletion fronts that invade the bulk of the drop from the contact line and localise the accumulation of stresses. Finally, finite element simulations are used to calculate the stress profiles in poroelastic drops with large contact
angles. We find that many of the conclusions obtained from the reduced model
still apply.

The rest of this paper is organised as follows. In Section \ref{sec:model}, details of the problem formulation and non-dimensionalisation are provided.  The governing equations are asymptotically reduced in Section~\ref{sec:drop_reduction}.  
The results are discussed in Section~\ref{sec:results}, while Section~\ref{sec:conclusion} is devoted to the concluding remarks.

\section{Problem formulation}
\label{sec:model}

We consider a drop of fluid consisting of a volatile solvent (e.g.\ water) and a non-volatile colloidal component (e.g.\ nanoparticles) that dries on a horizontal non-deformable solid substrate. We envision the colloids as having formed a poroelastic solid from which evaporation occurs. The contact line is assumed to remain pinned, which is in agreement with experiments, and the time-dependent contact angle $\varphi(t)$ is assumed to be small. The characteristic height and lateral extent (e.g.\ the radius) of the drop are denoted by $H$ and $R$, respectively, where $H / R \ll 1$.
We work within the framework of nonlinear poroelasticity~\cite{macminn2016} to account for large deformations during drying. We also assume that the drop remains bonded to the underlying substrate and thus neglect delamination processes that can potentially occur.

\subsection{Kinematics}

The governing equations are formulated in terms of Eulerian
coordinates $\vec{x} = x_i \vec{e}_i$
associated with the current (deformed) configuration, where $\vec{e}_i$ are
Cartesian basis vectors and summation over repeated indices is implied.
We let $\vec{X} = X_i \vec{e}_i$ denote
Lagrangian coordinates associated with the initial (undeformed) configuration
of the drop.
During drying, the solid element originally located
at $\vec{X}$ is displaced to $\vec{x}$, thereby generating a displacement
$\as = \vec{x} - \vec{X}(\vec{x},t)$. The deformation gradient
tensor $\tens{F}$ and its determinant $J = \det \tens{F}$ describe the distortion 
and volumetric changes of material elements, respectively. In Eulerian
coordinates, the deformation gradient tensor
is most readily expressed in terms of its inverse as 
\begin{align}
  \tens{F}^{-1} = \nabla \vec{X} = \tens{I} - \nabla \as,
  \label{eqn:F_inv}
\end{align}
where $\nabla$ is the spatial gradient taken with respect to the
Eulerian coordinates $\vec{x}$. We adopt the convention that
$\nabla \as = (\pdf{u_{i}^s}{x_j})\,\vec{e}_i\otimes\vec{e}_j$.
The velocity of the fluid and solid are written as
$\vf$ and $\vs$, respectively. In Eulerian formulations of nonlinear elasticity, the rate of change of displacement is linked to the velocity by the relationship
\begin{align}
  \pd{\as}{t} + (\vs \cdot \nabla) \as = \vs,
  \label{eqn:a}
\end{align}
where $\as = {\bf 0}$ when $t = 0$.

\subsection{Balance laws}

The composition of the mixture is described by the volume fractions of fluid and solid, $\phif$ and $\phis$, respectively.
Conservation of liquid and solid yield
\subeq{
  \label{eqn:phi}
\begin{align}
  \pd{\phif}{t} + \nabla \cdot (\phif \vf) &= 0, \\
  \pd{\phis}{t} + \nabla \cdot (\phis \vs) &= 0.
\end{align}
}
In deriving \eqref{eqn:phi}, it has been assumed that the densities of the solid skeleton and the fluid are constant, that is, both phases are incompressible. Furthermore, it is assumed that no volume change occurs upon mixing and that material elements
only consist of solid and fluid, the latter of 
which leads to the condition
\begin{align}
  \phif + \phis &= 1.
                  \label{eqn:no_void}
\end{align}
Since the pore space is only occupied by fluid, the volume fraction of
fluid $\phif$ also represents the porosity of the solid. 
Due to the incompressibility of the fluid and solid, volumetric changes in material elements can only be due to imbibition or depletion of fluid within the pore space, leading to the relationship
\begin{align}
  J = \det \tens{F} = \frac{1 - \phif_0}{1 - \phif},  \label{eqn:J}
\end{align}
where $\phif_0$ represents the fluid fraction in the initial
undeformed configuration.  For simplicity, we assume
that $\phif_0$ is spatially uniform. 
The Jacobian determinant $J$ describes the
local contraction of the solid skeleton ($J < 1$) that occurs due to the loss of fluid from the pore space ($\phif < \phif_0$).  If the porosity $\phif$ remains spatially uniform during drying, then the Jacobian determinant can be written in terms of the total drop volume $V$ as $J(t) = V(t) / V(0)$.


The fluid within the pore space is assumed to be transported by pressure
gradients. Hence, we impose Darcy's law,
\begin{align}
  \phif (\vf - \vs) &= -\frac{k(\phif)}{\mu_f} \nabla p, \label{eqn:darcy}
\end{align}
where $k$ is the permeability of the solid skeleton, $\mu_f$ is the fluid viscosity, and $p$ is the pressure.  The contraction of the solid matrix during drying will reduce the pore size and hence decrease the permeability.  Deformation-driven changes in the permeability are captured through its dependence on the porosity.  In particular, we adopt a normalised Kozeny--Carmen law~\cite{kozeny1927, carman1937} for the permeability given by
\begin{align}
    \frac{k(\phif)}{k_0} = \frac{(1 - \phif_0)^2}{(\phif_0)^3}\frac{(\phif)^3}{(1-\phif)^2},
\end{align}
where $k_0$ is the permeability of the initial configuration.

Conservation of momentum for the two-phase mixture yields
\begin{align}
    \nabla \cdot \vec{\sigma} &= \nabla p, \label{eqn:mo_mom}
\end{align}
where $\vec{\sigma}$ is the effective (Terzaghi) elastic stress tensor of the solid~\cite{coussy2004}, which commonly appears in soil mechanics~\cite{biot1941, terzaghi1936}. The solid skeleton is assumed to be isotropic and obey a neo-Hookean
equation of state. The elastic component of the stress tensor can be written as
\begin{align}
\begin{split}
  \vec{\sigma} = \frac{\nu E}{(1 + \nu)(1-2\nu)}(J - 1)\tens{I} + \frac{E}{2(1+\nu)J}(\tens{B} - \tens{I}),
  \end{split}
  \label{eqn:sigma}
\end{align}
where $E$ is the Young's modulus, $\nu$ is Poisson's ratio (both assumed constant), $\tens{I}$ is the identity tensor, and
$\tens{B} = \tens{F} \tens{F}^T$ is the left Cauchy--Green tensor.
In the limit of small deformations, $\nabla \as \ll 1$, we find that $\tens{F} \sim \tens{I} + \nabla \as$, which implies that $\tens{B} \sim \tens{I} + \nabla \as + (\nabla \as)^T$ and $J = \det \tens{F} \sim 1 + \nabla \cdot \as$.
Hence, the stress-strain relation \eqref{eqn:sigma} reduces
to
\begin{align}
    \vec{\sigma} \sim &\frac{\nu E}{(1 + \nu)(1-2\nu)}(\nabla \cdot \as)\tens{I} \notag
    \\ &\quad + \frac{E}{2(1+\nu)}\,\left(\nabla \as + (\nabla \as)^T\right),
  \label{eqn:sigma_lin}
\end{align}
thus recovering linear elasticity.

It is convenient to decompose vector and tensor quantities into in-plane and
transverse components that are parallel and perpendicular to the substrate,
respectively. We let $x_3 = z$ and $X_3 = Z$
denote the transverse Eulerian and Lagrangian coordinates, respectively,
and let $\vec{e}_3 = \vec{e}_z$ be the corresponding basis vector. 
If $\vec{a} = a_i \vec{e}_i$ denotes an arbitrary vector, then we write
$\vec{a} = \vec{a}_{\parallel} + a_z \vec{e}_z$, where
$\vec{a}_{\parallel} = a_\alpha \vec{e}_\alpha$ is a vector of the 
in-plane components and $a_z = a_3$ is the transverse
component; here we adopt the convention that Greek indices are equal to 1 or 2.
Similarly, we introduce the in-plane gradient operator $\nx = \nabla - \ez\,\pdf{}{z}$. The symmetric elastic stress tensor $\tens{\sigma}$ is written
in terms of its in-plane components $\sxx = \sigma_{\alpha \beta}\vec{e}_\alpha \otimes \vec{e}_\beta$, transverse shear components $\sxz = \sigma_{\alpha 3} \vec{e}_\alpha$, and vertical component $\szz$ as 
$\tens{\sigma} = \sxx + \sxz \otimes \vec{e}_z + \vec{e}_z \otimes \sxz + \szz \vec{e}_z \otimes \vec{e}_z$. Similar decompositions will be used for other
tensorial quantities as well.

\subsection{Boundary and initial conditions}

We assume that the solid skeleton perfectly adheres to the rigid substrate, resulting
in a no-displacement condition
\begin{align}
  \as = {\bf 0}, \quad z = 0. \label{bc:adhesion}
\end{align}
In addition, the substrate is taken to be impermeable; therefore,
\begin{align}
  \vf \cdot \vec{e}_z = 0, \quad z = 0. \label{bc:no_flow}
\end{align}
The static contact line of the drop is denoted by the curve $\rcx$ and defined by the equation
\begin{align}
  h = 0, \quad \xx = \rcx.
\end{align}
The kinematic boundary conditions for the fluid and solid phase at the free surface are given by
\subeq{
\begin{alignat}{2}
  \rho_f \phif (\vf\cdot \vec{n} - v_n) &= \phif \q, &\quad z &= h(\xx,t), \label{eqn:kin_l}
  \\
  \rho_s \phis (\vs\cdot \vec{n} - v_n) &= 0, &\quad z &= h(\xx,t), \label{eqn:kin_s}
\end{alignat}}
where $\rho_f$ and $\rho_s$ are the densities of the fluid and solid, respectively; $\q = \q(\phif)$ is the evaporative mass flux which depends on the surface composition; and $v_n$ is the normal velocity of the free surface, defined as
\begin{align}
v_n = \frac{1}{(1 + |\nx h|^2)^{1/2}}\pd{h}{t}.
\end{align}
Continuity of stress at the drop surface is given by
\begin{align}
  \tens{\sigma}\cdot \vec{n} - p \vec{n} = \vec{0}, \quad z = h(\xx,t),
\end{align}
where the atmospheric pressure has been set to zero. Similar boundary
conditions on the stress have been used by other researchers
when modelling drying-induced stresses in colloidal suspensions \cite{bouchaudy2019, style2011}.
The initial conditions for the fluid fraction, displacement, and drop thickness are 
given by
\subeq{
\begin{align}
\phif(\vec{x},0) &= \phif_0, \\
\as(\vec{x},0) &= \vec{0}, \\ 
h(\xx,0) &= h_0(\xx).
\end{align}
}

The initial conditions can be placed in the
context of drying colloidal dispersions by connecting 
the quantities  $h_0$ and $\phif_0$ to the profile of the 
drop and the volume fraction of liquid at the point of gelation.  
The gel point depends on the nature of the colloids as well
as the evaporation conditions, as these control the possible
arrangements of particles (e.g.\ random close packing, face-centred cubic).
The experimental observation of gelation fronts implies that different regions of the drop undergo the sol-gel transition at different times, making it difficult to define a profile for $h_0$. However, some insights 
can be obtained from the shape of the solid deposit that remains
on the substrate when drying is complete. 
Anyfantakis \etal\cite{anyfantakis2017} examined the drying of 
aqueous drops containing silica nanoparticles;  by increasing
the hydrophobicity of the nanoparticles, they observed
that the final deposit takes on a parabolic profile.  Therefore, the colloids likely remained uniformly dispersed during drying, which would have resulted in
a homogeneous gelation when the drop had a parabolic profile. 
However, by decreasing the hydrophobicity of the particles, the
deposit had a non-monotonic profile. 
Given the different profiles that might 
arise during drying, we will treat $h_0$ as a
parameter with the aim of elucidating the role it plays
in determining the poromechanical response of the drop.

\subsection{Scaling and  non-dimensionalisation}
\label{sec:drop_scaling}

We scale spatial quantities according to
$\xx \sim R$, $z \sim H$, $h \sim H$, and define
$\epsilon = H / R \ll 1$. 
For the liquid, we choose the usual lubrication scales for the velocity, $\vfx \sim V$ and $\vfz \sim \epsilon V$, where the velocity scale $V$ will be defined below. We use an advective time scale given by $t \sim R / V$.

To facilitate identifying appropriate scales for the solid,
we assume that the linear stress-strain relation given by
\eqref{eqn:sigma_lin} applies. Balancing $\vsz$ with $\pdf{h}{t}$ in the kinematic boundary condition \eqref{eqn:kin_s} implies that $\vsz \sim \epsilon V$. Moreover, balancing $\pdf{\asz}{t}$ with $\vsz$ in the vertical component of \eqref{eqn:a} gives a scale for the vertical displacement: $\asz \sim H$. A scale for the vertical normal stress can then be obtained as $\sigma_{zz} \sim E\, \pdf{\asz}{z} \sim E$. We postulate that large horizontal contractions of the solid skeleton are prohibited by its adhesion to the substrate. Therefore, the removal of solvent will drive a predominantly vertical contraction of the solid skeleton. The elastic stress that is generated by this vertical contraction, $\sigma_{zz}$, must be balanced by the pressure, $p$, resulting in $p \sim \sigma_{zz} \sim E$. A scale for the horizontal displacements can be obtained through a consideration of the horizontal momentum balance for the mixture. As in lubrication theory, we expect the horizontal pressure gradient to generate a shear stress. Therefore, we balance $\pdf{\sxz}{z}$ with $\nx p$ which implies that $\sxz \sim \epsilon E$. The shear stress also scales like $\sxz \sim E\,\pdf{\asx}{z}$; thus, we find that $\pdf{\asx}{z} \sim \epsilon$. In light of the no-slip condition for the solid, this balance implies that $\asx \sim \epsilon H$ and hence $\vsx \sim \epsilon^2 V$. The in-plane components of the stress tensor can be scaled as $\sxx \sim E$.
Finally, the velocity scale is determined from the horizontal component of Darcy's law, which gives $V = (k_0/\mu_f)(E/R)$.

Under this scaling, the non-dimensional displacement gradient is
\begin{align}
  \nabla \as = \epsilon^2 \nx \asx + \epsilon\left(\pd{\asx}{z}\otimes \ez + \ez \otimes \nx \asz\right) \notag\\ 
  \quad + \pd{\asz}{z}\ez \otimes \ez,
  \label{nd:grad_u}
\end{align}
which shows that in-plane and shear strains will be small. The non-dimensional
form of the elastic stress tensor is
$\tens{\sigma} = \sxx + \szz \vec{e}_z \otimes \vec{e}_z + \epsilon(\sxz \otimes \vec{e}_z + \vec{e}_z \otimes \sxz)$.

\subsubsection{Non-dimensional bulk equations}
\label{sec:nd_drop_bulk}
The rescaled conservation equations for the volume fractions are given by
\subeq{
  \begin{align}
    \pd{\phif}{t} + \nx\cdot\left(\phif \vfx\right) + \pd{}{z}\left(\phif \vfz \right) &= 0, \label{rs:cons_fluid} \\
    \pd{\phis}{t} + \epsilon^2\nx\cdot\left(\phis \vsx\right) + \pd{}{z}\left(\phis \vsz \right) &= 0. \label{rs:cons_solid}
  \end{align}
}
Thus, fluid is transported in both the horizontal and vertical directions.
The solid, however, is predominantly transported in the vertical direction,
suggesting that a uniaxial mode of deformation occurs. 
The components of the solid velocity can be obtained from
\subeq{\label{rs:vs}
  \begin{align}
    \pd{\asx}{t} + \epsilon^2 \vsx \cdot \nx \asx + \vsz \pd{\asx}{z} = \vsx, \\
    \pd{\asz}{t} + \epsilon^2 \vsx \cdot \nx \asz + \vsz \pd{\asz}{z} = \vsz.
  \end{align}
}
Darcy's law \eqref{eqn:darcy} can be written in component form as
\subeq{
\begin{align}
  \phif (\vfx - \epsilon^2 \vsx) &= -k(\phif)\nx p, \label{rs:mom_f_x} \\
  \epsilon^2 \phif (\vfz - \vsz) &= -k(\phif)\pd{p}{z}, \label{rs:mom_f_z}
\end{align}
}
which shows that vertical gradients in the pressure will be weak.  The in-plane motion of the solid skeleton plays a sub-dominant role in \eqref{rs:mom_f_x} and can be neglected.  Consequently, the problems describing the in-plane transport of solid and fluid decouple. The momentum balance for the mixture \eqref{eqn:mo_mom} is
\subeq{
\begin{align}
  \nx \cdot \sxx + \pd{\sxz}{z} &= \nx p, \label{rs:mom_x} \\
  \epsilon^2 \nx \cdot \sxz + \pd{\szz}{z} &= \pd{p}{z}. \label{rs:mom_z}
\end{align}
}
The lubrication approximation for poroelastic solids therefore leads to a different stress balance than for viscous fluids by bringing the in-plane stresses $\sxx$ and
the vertical normal stress $\szz$ into the leading-order problem.

\subsubsection{Non-dimensional boundary and initial conditions}
\label{sec:nd_drop_bc}

The non-dimensional adhesion and no-flux conditions are the same as in
\eqref{bc:adhesion} and \eqref{bc:no_flow}.
The kinematic boundary conditions become
\subeq{
\begin{alignat}{2}
  \pd{h}{t} + \vfx \cdot \nx h - \vfz = -\mathcal{Q} q(\phif)\mathcal{A}, \,\,\,\, z = h(\xx,t), \label{bc:kin_f}\\
 \pd{h}{t} + \epsilon^2 \vsx \cdot \nx h - \vsz = 0, \quad z = h(\xx,t),
\end{alignat}}
where $\mathcal{Q} = q_0 / (\rho_f V \epsilon)$, $q_0 = q_e(\phif_0)$ is the initial evaporative mass flux, and $\mathcal{A} = \left(1 + \epsilon^2|\nx h|^2\right)^{1/2}$. The non-dimensional parameter $\mathcal{Q}$
plays the role of a P\'eclet number by characterising the relative rate of evaporation to bulk fluid transport. 
The stress balances at the free surface are given by
\subeq{
  \begin{alignat}{2}
    -\sxx \cdot \nx h + \sxz + p\nx h &= \vec{0}, &\quad z &= h(\xx,t),
    \label{rs:bc:tan_stress}
    \\
    -\epsilon^2 \sxz \cdot \nx h + \szz - p &= 0, &\quad z &= h(\xx,t).
    \label{rs:bc:norm_stress}
\end{alignat}
}
The initial conditions for the drop profile, porosity, and displacements are
$h = h_0(\xx)$, $\phif = \phif_0$, $\asx = \vec{0}$, and $\asz = 0$ when $t = 0$.

\subsection{Parameter estimation}
\label{sec:params}

Giorgiutti--Dauphin\'e and Pauchard \cite{giorgiutti2014} conducted experiments
on colloidal drops consisting of silica nanoparticles in water.
They reported values of
$k_0 \sim 10^{-19}$~m\unit{2}, $E \sim 1$~GPa, $\mu_f \sim 10^{-3}$~Pa~s,
and
$R \sim 1$~mm.  The initial contact angle $\varphi_0$ ranged from
$30^\circ$ to $40^\circ$, leading to values of 
$\epsilon \sim \varphi_0$ in the range of 0.5 to 0.7.  
The evaporation velocity, $V_e \sim q_0 / \rho_f$,
can be inferred from their measurements of the cracking time
and is found to be
roughly $10^{-9}$~m~s\unit{-1}.  A conservative estimate
of the P\'eclet number based on a value of $\epsilon = 0.1$ 
is then $\mathcal{Q} \sim 10^{-3}$.
Osman \etal\cite{osman2020} reported similar parameter values
for their experiments: $k_0 \sim 10^{-20}$~m\unit{2},
$V_e \sim 10^{-8}$~m~s\unit{-1}, $\mu_f \sim 10^{-3}$~Pa~s, 
and $R \sim 1$~mm.  The Young's modulus and contact angles were not measured.
However, since their colloidal dispersions were also based on 
silica nanoparticles,
we estimate that $E \sim 1$~GPa. The P\'eclet number
can be parameterised in terms of the initial contact angle
as $\mathcal{Q} \sim 10^{-3}\, \varphi_0^{-1}$ and is expected to be small.
Finally, in the case of drying blood drops, 
Sobac and Brutin~\cite{sobac2014} reported that
$R = 4.3$~mm, $V_e \simeq 9\cdot 10^{-8}$~m~s\unit{-1}, and
$\varphi_0 = 15^\circ$.  Moreover, they estimated
that the diffusivity of fluid through the poroelastic
solid was roughly $D_w \simeq 3\cdot 10^{-8}$~m\unit{2}~s\unit{-1},
which leads to a velocity scale of 
$V \sim D_w / R \simeq 7\cdot 10^{-6}$~m~s\unit{-1}.  
The corresponding P\'eclet number is $\mathcal{Q} \sim 0.05$. 

\section{Asymptotic reduction} 
\label{sec:drop_reduction}

The dimensionless model is asymptotically reduced by taking the limit as $\epsilon \to 0$.
The reduction can be decomposed into two main steps.  First, the mechanical problem is solved
in terms of the fluid fraction. Second, the transport problems for the fluid and solid are
simplified and then combined into a thin-film-like equation for the drop thickness.  

The reduction of the mechanical problem begins with a consideration of the rescaled displacement
gradient \eqref{nd:grad_u} and the deformation gradient tensor \eqref{eqn:F_inv}.
By taking $\epsilon \to 0$ in \eqref{nd:grad_u} and substituting the result in
\eqref{eqn:F_inv}, the leading-order contribution 
to the deformation gradient tensor 
can be written as $\tens{F} = \Ixx + J \vec{e}_z \otimes \vec{e}_z$, where
$\Ixx = \vec{e}_\alpha \otimes \vec{e}_\alpha$ is the in-plane identity tensor
and
\begin{align}
  J = \det \tens{F} = \left(1 - \pd{\asz}{z}\right)^{-1}.
  \label{red:J_c}
\end{align}
The Jacobian determinant $J$ in \eqref{red:J_c} must also satisfy \eqref{eqn:J}. 
The asymptotic reduction of the deformation gradient tensor $\tens{F}$
shows that, to leading order, the drop undergoes uniaxial
deformation along the vertical direction. 
The leading-order components of the elastic stress tensor are
\subeq{\label{red:sigma}
\begin{align}
  \sxx &= \frac{\nu}{(1+\nu)(1-2\nu)}(J - 1)\Ixx, \label{red:sxx}\\
  \sxz &= \frac{1}{2(1+\nu)}\left(J \pd{\asx}{z} + \nx \asz\right),
         \label{red:sxz}\\
  \szz &= \frac{1}{1+\nu}\left[\frac{1}{2}(J - J^{-1}) + \frac{\nu}{1-2\nu}(J-1)\right]. \label{red:szz}
\end{align}}
By integrating the $O(1)$ contributions to the vertical stress balance
\eqref{rs:mom_z},
we observe that the pressure is equal to the vertical normal stress,
\begin{align}
  p = \szz = \frac{1}{1+\nu}\left[\frac{1}{2}(J - J^{-1}) + \frac{\nu}{1-2\nu}(J-1)\right].
  \label{eqn:p_J}
\end{align}
Taking $\epsilon \to 0$ in \eqref{rs:mom_f_z} shows that the pressure $p$ is independent of $z$ to leading order.  Consequently, we deduce that $\szz$, $J$, and hence $\phif$ are also independent of $z$.
The horizontal stress balance \eqref{rs:mom_x} can now
be integrated and the stress-free condition
\eqref{rs:bc:tan_stress} imposed to find
\begin{align}
  \sxz = \frac{1}{2(1+\nu)}\nx\left((h-z)(J^{-1} - J)\right).
  \label{red:sxz_sol}
\end{align}
Equating \eqref{red:sxz} and \eqref{red:sxz_sol}
leads to a differential equation for the in-plane
components of the solid displacement $\asx$. Furthermore, \eqref{red:J_c}
provides an equation
for the vertical displacement $\asz$. Upon solving these equations and imposing
$\as = \vec{0}$ at $z = 0$, we
find that the displacements are given by
\subeq{
  \label{red:disp}
\begin{align}
  \asx &= \frac{1}{2}\,z^2\,\nx(\log J) + z J^{-1} \nx\left(h (J^{-1} - J)\right),
         \label{red:disp_x}
  \\
  \asz &= (1 - J^{-1}) z.
         \label{red:disp_z}
\end{align}}
At this point, the mechanical problem has been completely solved in terms
of the Jacobian determinant $J$, which is linked to the fluid
fraction via \eqref{eqn:J}. 

At leading order, the conservation law for the fluid \eqref{rs:cons_fluid} becomes
\begin{align}
  \pd{\phif}{t} + \nx\cdot(\phif \vfx) + \phif \pd{\vfz}{z} = 0.
  \label{red:phi_f}
\end{align}
Integrating \eqref{red:phi_f} across the thickness of the drop and using 
the impermeability condition \eqref{bc:no_flow} and
the kinematic boundary condition \eqref{bc:kin_f} leads to
\begin{align}
  \pd{}{t}(h \phif) + \nx\cdot (h \phif \vfx) = -\mathcal{Q} \phif q(\phif).
  \label{eqn:tf_f}
\end{align}
Similarly, from the conservation of solid \eqref{rs:cons_solid}, we find that
\begin{align}
  \pd{}{t}(h \phis) = 0
  \label{eqn:tf_s}
\end{align}
to leading order. By integrating \eqref{eqn:tf_s} and using the definition of
$J$ from \eqref{eqn:J}, we obtain
\begin{align}
  h = J h_0.
  \label{red:h_phi}
\end{align}
Equation \eqref{red:h_phi} reflects the uniaxial mode of deformation that the drop
experiences and states that volumetric changes in material elements
can only be accommodated through variations
in the film thickness $h$. 
Adding \eqref{eqn:tf_f} and \eqref{eqn:tf_s} and using Darcy's law \eqref{rs:mom_f_x} gives
\begin{align}
  \pd{h}{t} = \nx \cdot \left[k(\phif) h \nx p\right] - \mathcal{Q} \phif q(\phif).
  \label{eqn:thin_film_p}
\end{align}
By using \eqref{eqn:p_J} and \eqref{red:h_phi} to write $p = \szz(J)$ and $J = h / h_0$, respectively, \eqref{eqn:thin_film_p} can be formulated
as a thin-film-like equation
\subeq{
  \label{eqn:thin_film_sys}
\begin{align}
  \pd{h}{t} = \nx \cdot \left[k(\phif) h \szz'\left(\frac{h}{h_0}\right) \nx \left(\frac{h}{h_0}\right)\right] 
  - \mathcal{Q} \phif q(\phif),
    \label{eqn:thin_film}
\end{align}
where $\szz'(J) = \d \szz(J) / \d J$ and the solvent fraction is given by
$\phif = 1 - (1 - \phi_{f,0})(h_0/h)$. The thin-film equation \eqref{eqn:thin_film} can be solved using the boundary and initial conditions
\begin{alignat}{2}
  h &= 0, &\quad \xx &= \rcx; \\
  h &= h_0(\xx), &\quad t &= 0.
\end{alignat}}
Once the drop thickness $h$ is calculated, the Jacobian determinant
$J$ can be computed from \eqref{red:h_phi} and used to evaluate
the elastic stresses and displacements given by
\eqref{red:sigma}--\eqref{red:disp}. 


\subsection{The slow-evaporation limit}

The parameter estimates from Sec.~\ref{sec:params} indicate that the
P\'eclet number $\mathcal{Q}$ is typically small, implying that fluid loss due to
evaporation is slow relative to the rate at which fluid is replenished by
bulk transport.  This separation of time scales can be used to
further reduce the model.  By rescaling time as
$t = \mathcal{Q}^{-1} \tau$ and taking $\mathcal{Q} \to 0$, we can deduce from
\eqref{eqn:thin_film} and \eqref{red:h_phi} that $h / h_0$ and hence $J$
must be spatially uniform. This permits the film thickness to be written as
$h(\xx,\tau) = J(\tau) h_0(\xx)$. Consequently,
the fluid fraction $\phif$ must also be independent of space in order
to satisfy the incompressibility condition \eqref{eqn:J}. 
To determine the time dependence of $J$, we integrate \eqref{eqn:thin_film}
over the contact surface to obtain
\begin{align}
  \td{J}{\tau} = -\frac{A_0 \phif q(\phif)}{V_0},
  \label{eqn:dJdt}
\end{align}
where $A_0$ and $V_0$ are the (non-dimensional) 
area of the contact surface and the initial
volume of the drop, respectively, and $\phif = 1 - (1 - \phi_{f,0})/J$.
Equation \eqref{eqn:dJdt} can be recast
into a differential equation for the volume of the drop $V(\tau)$ using
the relation $J(\tau) = V(\tau) / V_0$.

\section{The poromechanics of drying}
\label{sec:results} 

The solutions of the asymptotically reduced model provide new
insights into the poromechanics of drying drops.  We first analyse
and interpret the solutions for the stress.
We then explore the mechanics of drying in the
limit of slow evaporation, corresponding to vanishingly small
P\'eclet numbers $\mathcal{Q}$.  Numerical simulations are used
to study the dynamics for moderate evaporation rates characterised by
P\'eclet numbers that are $O(1)$ in size. Finally, we use finite element
simulations to examine the stresses that arise when the contact angle
is not small.

\subsection{Analysis of drying-induced stresses}

A key feature of the asymptotic reduction is that it allows for  
a straightforward determination and interpretation of the
stresses that are generated during drying.  The total
stress within the poroelastic drop is characterised
by the Cauchy stress tensor $\tens{T} = \tens{\sigma} - p \tens{I}$
and can therefore be decomposed into an
elastic stress associated with deformations of the solid skeleton
and an isotropic contribution arising from the fluid. 
Since drying leads to a loss of volume ($J < 1$),
we see from \eqref{red:sxx} that in-plane elastic stresses 
$\sxx$ are \emph{compressive}.
The origin of these compressive stresses can be understood by 
drawing on the analogy between the drying-induced contraction of the 
solid skeleton and the vertical compression of a slab of elastic material. 
Due to the Poisson effect, vertical compression of a slab will drive
a lateral (or in-plane) expansion. However,
if the slab is bonded to a substrate, then 
lateral expansion is constrained
and a compressive stress is generated to resist lateral deformation.
An examination of the total in-plane stresses, defined by
\begin{align}
  \Txx = \sxx - p\Ixx = \frac{1}{2(1 + \nu)}(J^{-1} - J)\Ixx,
  \label{eqn:Txx}
\end{align}
reveals they are \emph{tensile}
due to the negative pressure counteracting the elastic stresses. 
The combination of a tensile total stress and a compressive elastic stress is consistent with the findings of Bouchaudy and Salmon \cite{bouchaudy2019}, who report
similar mechanics in a one-dimensional setting.  

The generation of tensile in-plane
stresses leads to a mechanism for fracture, which is commonly observed during the 
drying of complex colloidal suspensions.  However, the isotropic form of the
in-plane stress tensor \eqref{eqn:Txx}
prohibits the leading-order problem from providing any information about the 
orientation of nucleated fractures, which we expect to be perpendicular to the directions of maximal stress.  Any mechanism that could select a 
preferential direction for fracture must therefore manifest
in higher-order contributions to the stress tensor and, as a result, be relatively weak.  


Drying-induced stresses can trigger the delamination of the drop
from rigid substrates, in which case
knowledge of the traction
exerted by the drop on the substrate is crucial.
The non-dimensional traction is defined as
$\TT = \tens{T}|_{z=0}\cdot \vec{e}_z$.
The in-plane components of the traction,
which are generated from elastic shear stresses,
are readily computed from \eqref{red:sxz_sol} and
found to be
\subeq{\label{drop:TT}
\begin{align}
  \TTx = \frac{1}{2(1+\nu)}\nx\left(h(J^{-1} - J)\right).
  \label{drop:Tx}
\end{align}
To determine the leading-order component of the vertical traction $\TTz$, we
integrate \eqref{rs:mom_z} across the thickness of the drop and
impose the stress-free
condition \eqref{rs:bc:norm_stress} to obtain
\begin{align}
  \TTz = \frac{\epsilon^2}{4(1+\nu)}\nx^2\left(h^2(J^{-1} - J)\right).
  \label{drop:Tz}
\end{align}}
The vertical traction \eqref{drop:Tz} can be
interpreted as the adhesive stress required for the drop to remain
bonded to the substrate during drying. Positive and negative values of
$\TTz$ imply that the drop is pulling upwards and pushing downwards
on the substrate, respectively. 
Due to the prefactor of $\epsilon^2 \sim \varphi_0^2$ 
appearing in \eqref{drop:Tz},
thinner drops with
smaller contact angles will be less prone to delamination, 
provided this occurs
once $\TTz$ exceeds a critical threshold.  To the best of our knowledge,
the contact-angle dependence of delamination has not been
investigated experimentally.

\subsection{Mechanics in the slow-evaporation limit}

Significant insight into the mechanics of drying 
can be obtained by considering the slow-evaporation limit, as the spatial uniformity of the Jacobian determinant
$J$ enables the asymptotic solutions to be greatly simplified.  
Due to the monotonic decrease in the drop volume $V$ in time, 
we can use $J(t) = V(t) / V_0$ as a proxy for time, 
where $J$ decreases from $J = 1$ when $t = 0$ to a steady-state value
of $J = J_\infty < 1$ as $t \to \infty$.  For simplicity, we focus
on the case of axisymmetric drops with circular contact lines. 

The radial displacement can be computed from \eqref{red:disp_x} and is found to be
\begin{align}
\asr = z (J^{-1} - J) \td{h_0}{r}.
\label{slow:asr}
\end{align}
Consequently, the radial and orthoradial motion of the solid skeleton is controlled by the initial geometry of the drop, which is encoded in the functional
form of $h_0$. In locations where the initial profile has
a negative slope, $\d h_0 / \d r < 0$, solid elements are
displaced towards the drop centre ($\asr < 0$) and undergo orthoradial compression ($\asr / r < 0$).  However, if the initial profile has
a positive slope, $\d h_0 / \d r > 0$, then solid elements are displaced
towards the contact line ($\asr > 0$) and undergo orthoradial
expansion ($\asr / r > 0$).  Similarly, the initial curvature of the solid skeleton, 
$\d^2 h_0 / \d r^2$, controls the mode of radial deformation.  
Solid elements experience a radial
compression ($\pdf{\asr}{r} < 0$) if the curvature is negative and a radial expansion ($\pdf{\asr}{r} > 0$) if the
curvature is positive. 

The compressive and extensional modes of radial and orthoradial deformation
lead to small differences in the radial and hoop stresses. Although
small, these differences
can establish a preferential direction for nucleated fractures.
By calculating the higher-order terms in the elastic stress tensor, 
we find that the difference between the radial
and hoop stress can be expressed as
$\mathsf{T}_{rr} - \mathsf{T}_{\theta \theta} = \sigma_{rr} - \sigma_{\theta\theta} = \epsilon^2 (1+\nu)^{-1} \mathcal{S}(r,z)$, where 
\begin{align}
\mathcal{S} = \pd{\asr}{r} - \frac{\asr}{r}
+ J \pd{\asr}{z}\pd{\asz}{r} + \frac{J^2}{2}\left(\pd{\asr}{z}\right)^2.
\label{eqn:S}
\end{align}
The first two terms on the right-hand side of \eqref{eqn:S} capture the competition
between linear radial and orthoradial strains, and they would be present if the model had been formulated in
terms of linear elasticity.  
The final two terms on the right-hand side of  \eqref{eqn:S} arise from
geometric nonlinearities associated with
finite strains. 
Substituting the expression for the radial displacement \eqref{slow:asr} and the vertical displacement
\eqref{red:disp_z} into \eqref{eqn:S} leads to
\begin{align}
\mathcal{S} = (J^{-1} - J) z 
\left(\tdd{h_0}{r}-\frac{1}{r}\td{h_0}{r}\right)
+ \frac{(J^2 - 1)^2}{2}\left(\td{h_0}{r}\right)^2.
\label{eqn:S0}
\end{align}
Using \eqref{eqn:S0}, it is straightforward to explore how
different drop profiles $h_0$ affect the competition between
radial and orthoradial stress generation. 

Near the contact line, the initial profile of the drop can be 
locally represented
as a linear function with negative gradient.  The resulting value 
of $\mathcal{S}$ will be positive,  indicating that
the radial stress dominates the hoop stress.
Consequently, fractures will have a slight preference to
align with (be parallel to) the orthoradial direction.  Due
to $\mathsf{T}_{rr} - \mathsf{T}_{\theta \theta}$
being proportional to $\epsilon^2 \sim \varphi_0^2$,
the strength of the
orthoradial alignment should increase with the initial
contact angle.  Observations of similar qualitative trends were made in the experimental work of 
Carle and Brutin~\cite[Fig.~4]{carle2013}
where increases in the initial contact angle
led to increasingly prominent orthoradial
fractures at the contact line.

Understanding the competition between the radial
and hoop stresses away from the contact line requires
specific knowledge of the initial profile
of the poroelastic drop. Parabolic profiles for
$h_0$ represent a special
case and lead to the first bracketed term in \eqref{eqn:S0} vanishing,
implying that the radial and orthoradial strains are
identical in the limit of linear elasticity and can only
be distinguished by considering a nonlinear theory. 
Although the value of $\mathcal{S}$ is always positive for parabolic drops,
the prefactor of the final term will be small and thus 
any preferential orientation of fractures will be very weak. 
The drying experiments by 
Anyfantakis \etal\cite{anyfantakis2017} support this
prediction: the solid deposits that
were parabolic in shape were patterned by disordered
fractures with no clear orientation.  
In this case, the parabolic profile is likely a result of the drops
remaining homogeneous during drying.

\begin{figure}
\centering
\includegraphics[width=0.9\columnwidth]{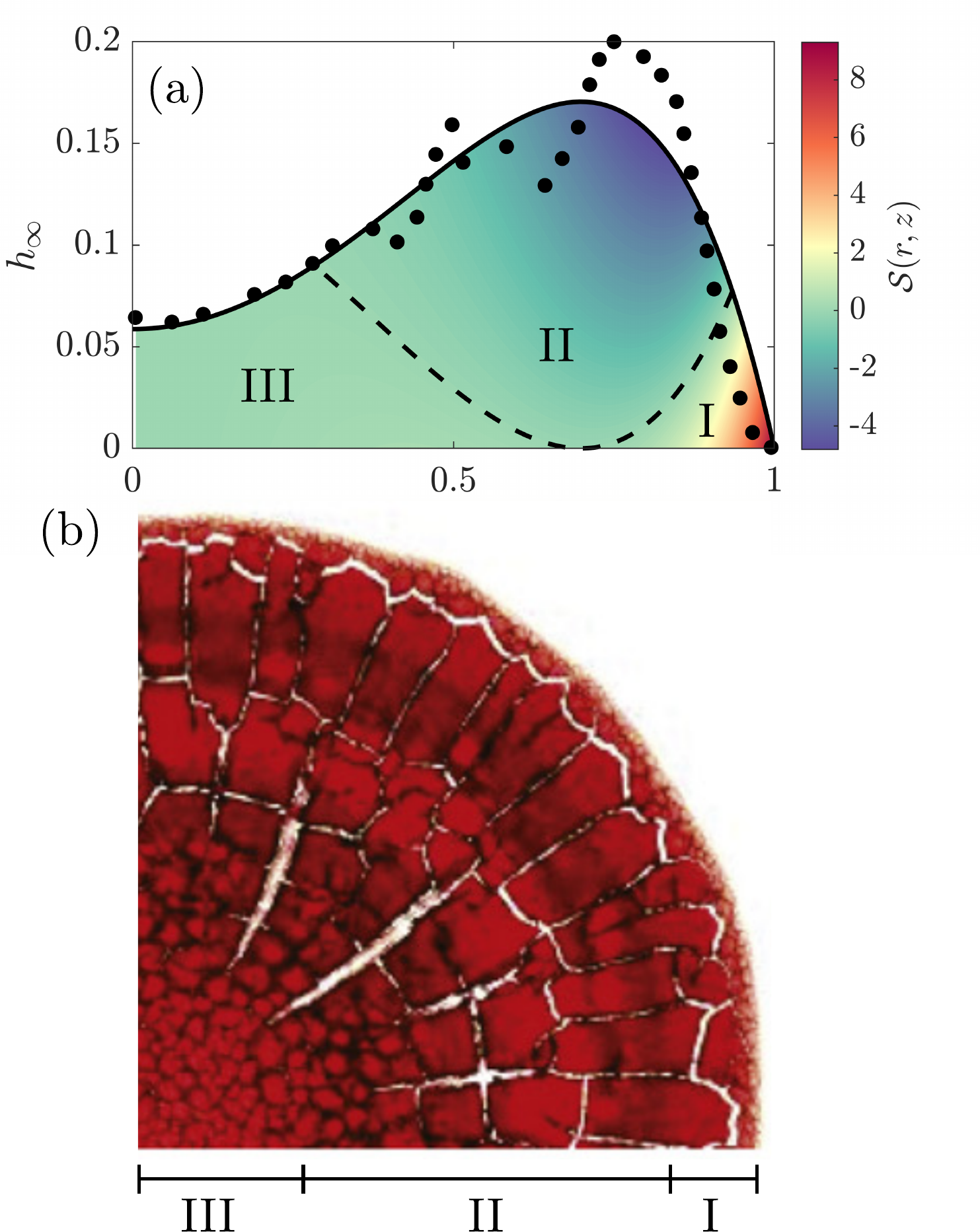}
\caption{The alignment of fractures is driven by the deposit profile.
(a) The steady-state profile of a dried
blood drop.  Circles represent experimental
data from Sobac and Brutin~\cite{sobac2014} and the solid line is a polynomial fit.  The superimposed heatmap illustrates the difference between the radial and hoop stresses $\mathcal{S} = \epsilon^{-2}(1+\nu)(\sigma_{rr} - \sigma_{\theta\theta})$.  The black dashed line is the $\mathcal{S} = 0$ level set. The three regions correspond to (I) a dominant radial stress (orthoradial fractures), (II) a dominant orthoradial stress (radial fractures) and (III) similar stresses (non-oriented fractures).  (b) The deposit associated with
the profile in panel~(a), showing the three fracture patterns suggested by the asymptotic theory. This figure has been adapted from
Sobac and Brutin~\cite{sobac2014}.}
\label{fig:drop}
\end{figure}

We therefore postulate that the alignment of nucleated fractures is
due to the heterogeneous gelation of drops and the creation of
poroelastic skeletons with non-parabolic profiles.  Sobac and
Brutin~\cite{sobac2014} measured the profile of a dried deposit 
that was patterned by strongly aligned fractures. The deposit thickness was
non-monotonic and found to generally increase towards the contact line until
a maximum was reached, after which the thickness rapidly decreased to zero; 
see Fig.~\ref{fig:drop}~(a).  The dried deposit exhibited three
distinct fracture patterns: (I) near the contact line
there was a large orthoradial fracture; (II) away from the contact line
there was an annular region
where fractures were predominantly aligned with the radial direction; and
(III) there was a central region with non-oriented fractures.  
An image of the dried deposit showing the three fracture patterns is
provided in Fig.~\ref{fig:drop}~(b).

The three fracture patterns observed by Sobac and
Brutin~\cite{sobac2014} can be rationalised by the poroelastic model.
Doing so requires reconstructing
the profile of the drop at the point of gelation, which is 
achieved using the relation $J = h / h_0 = (1 - \phif_0) / (1 - \phif)$ and taking $\phif$ and $h$ to be the equilibrium fluid fraction $\phif_\infty$ and
drop profile $h_\infty$, respectively.  Solving for the initial 
profile $h_0$ gives
\begin{align}
h_0(r) \simeq \left(\frac{1 - \phif_\infty}{1 - \phif_0}\right) h_\infty(r).
\label{eqn:recon}
\end{align}
Sobac and Brutin~\cite{sobac2011} estimate that the first fractures occur when
the solid fraction is roughly 30\%.  Therefore, we take the gel point to
be $\phif_0 \simeq 0.7$.  In addition, the solid deposit is assumed to be
completely dry, $\phif_\infty \simeq 0$.  A smooth function for $h_\infty$
is obtained by fitting a polynomial to the experimentally measured profile,
resulting in the solid black curve shown in Fig.~\ref{fig:drop}~(a).
Using the reconstruction of $h_0$ provided by \eqref{eqn:recon}, we 
calculate the difference between the radial and hoop stress via
\eqref{eqn:S0} and plot the values of $\mathcal{S}$ as a heatmap in
Fig.~\ref{fig:drop} (a).  As predicted, there is a region near the
contact line where the radial stress dominates ($\mathcal{S} > 0$),
resulting in the orthoradial fracture associated with pattern (I).  
However, there is also an intermediate region centred about the
maximum of the deposit thickness where the hoop stress dominates ($\mathcal{S} < 0$) and the onset of the radially aligned fractures associated with pattern (II) is expected.  Finally, near the drop centre, the value of $\mathcal{S}$ is very close to zero, suggesting the emergence of non-oriented fractures observed in pattern (III).  

The model predicts that the appearance of multiple fracture
patterns will be a generic feature of poroelastic skeletons that
have a non-monotonic initial profile.  Near a maximum in the profile,
where $\d h_0 / \d r \simeq 0$ and $\d^2 h_0 / \d r^2 < 0$, the hoop stress
will dominate the radial stress ($\mathcal{S} < 0$), suggesting the emergence of radially aligned fractures.  Conversely, local minima in the
profile would lead to orthoradially aligned fractures.  
The presence of multiple maxima
and minima in the deposit profile shown in Fig.~\ref{fig:drop}~(a) 
could explain the sequential realignment of fractures that is seen
in Fig.~\ref{fig:drop}~(b). 

For slowly evaporating drops, the normal component of the traction
reduces to
\begin{align}
\TTz = \frac{\epsilon^2}{4(1 + \nu)} \frac{J^2 (J^{-1} - J)}{r}\td{}{r}\left[r \td{}{r}\left(h_0^2\right)\right].
\label{slow:Tz}
\end{align}
The competition between the decrease in the drop height, captured through $J^2$,
and the
increase in elastic stress, captured through $J^{-1} - J$, results
in a non-monotonic evolution of the traction that reaches 
a maximum value when $J = 2^{-1/2} \simeq 0.71$.  There are
two ramifications of this finite maximum.  Firstly, it
implies that delamination is not guaranteed to occur.
Secondly, if delamination does occur, then 
the propagating delamination front may not reach
the drop centre by the end of the drying process.
In fact, the drop only pulls upwards on the substrate
in locations where the curvature of $h_0^2$ is 
positive.  For drops with initially parabolic profiles,
$h_0 = 1 - r^2$, the traction 
is positive for $r > 2^{-1/2}$, which sets a 
theoretical maximum on the depth of delamination. 

Osman \etal\cite{osman2020} experimentally observed a
limited depth of delamination in drying colloidal drops.
Using a simple model, 
they argued that heterogeneous gelation leads to a poroelastic
`foot' developing at the contact line, the length of which
controls the depth of delamination.  Our complementary theory
predicts that even a fully gelled drop could only undergo a partial degree of delamination.  When combined,
these two theories suggest that the extent of delamination ultimately
arises from an intricate interplay between the horizontal growth of the poroelastic solid as well as its shape.


\subsection{Numerical simulations}

\begin{figure*}
  \centering
  \subfigure[]{\includegraphics[width=0.32\textwidth]{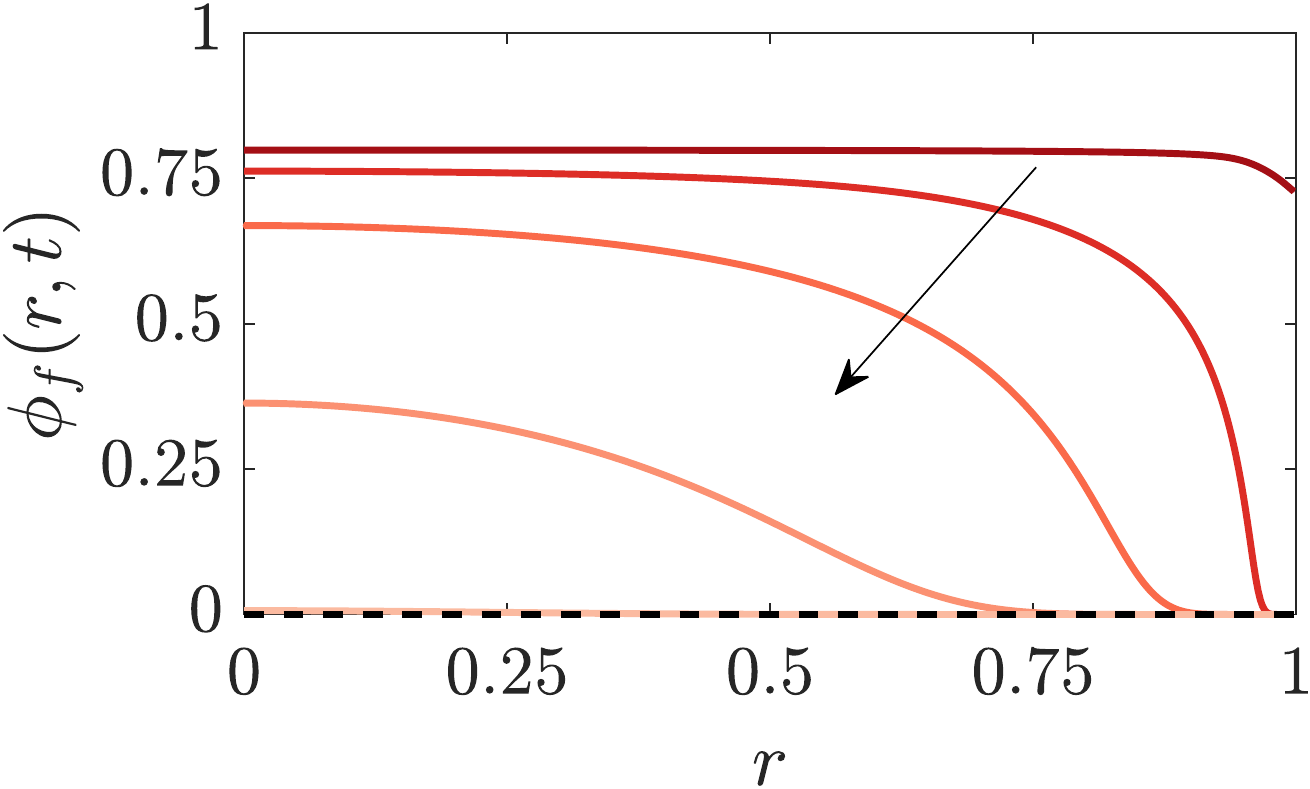}}
  \subfigure[]{\includegraphics[width=0.32\textwidth]{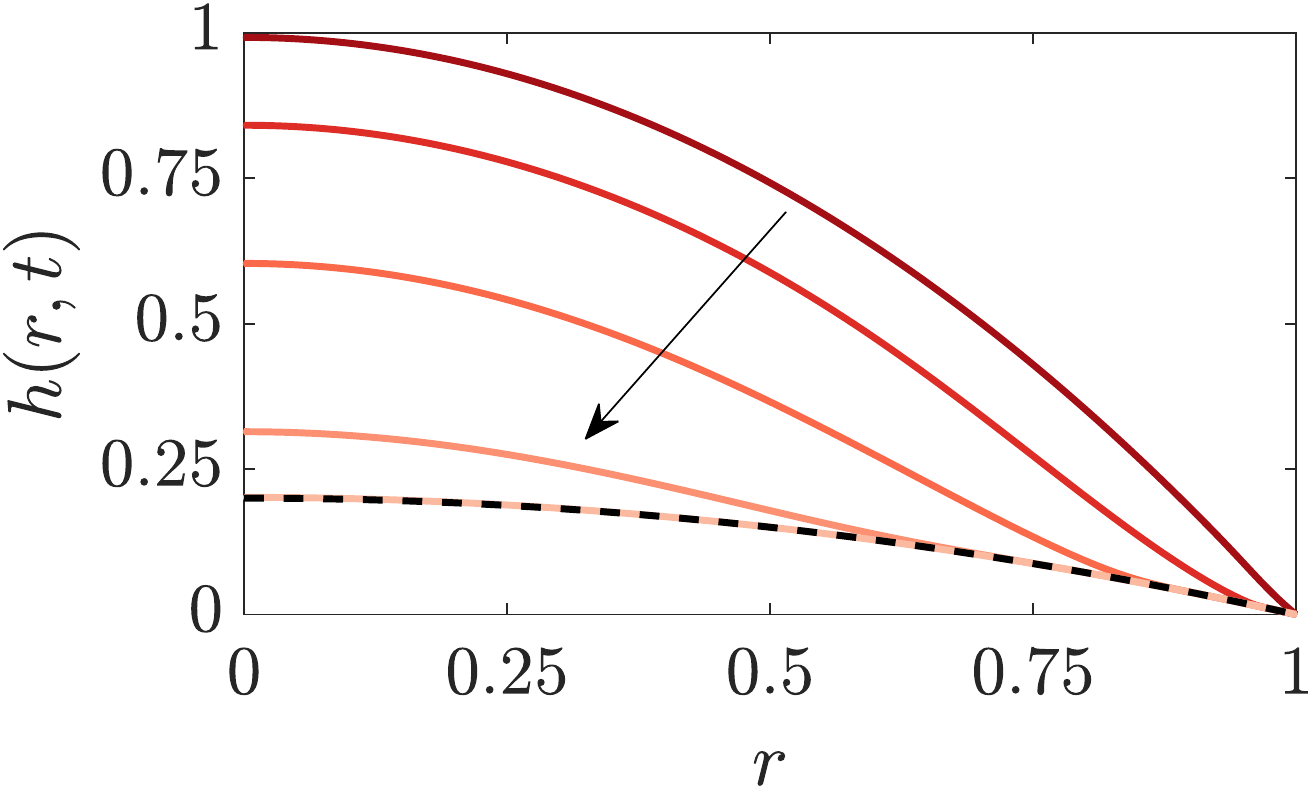}}
  \subfigure[]{\includegraphics[width=0.32\textwidth]{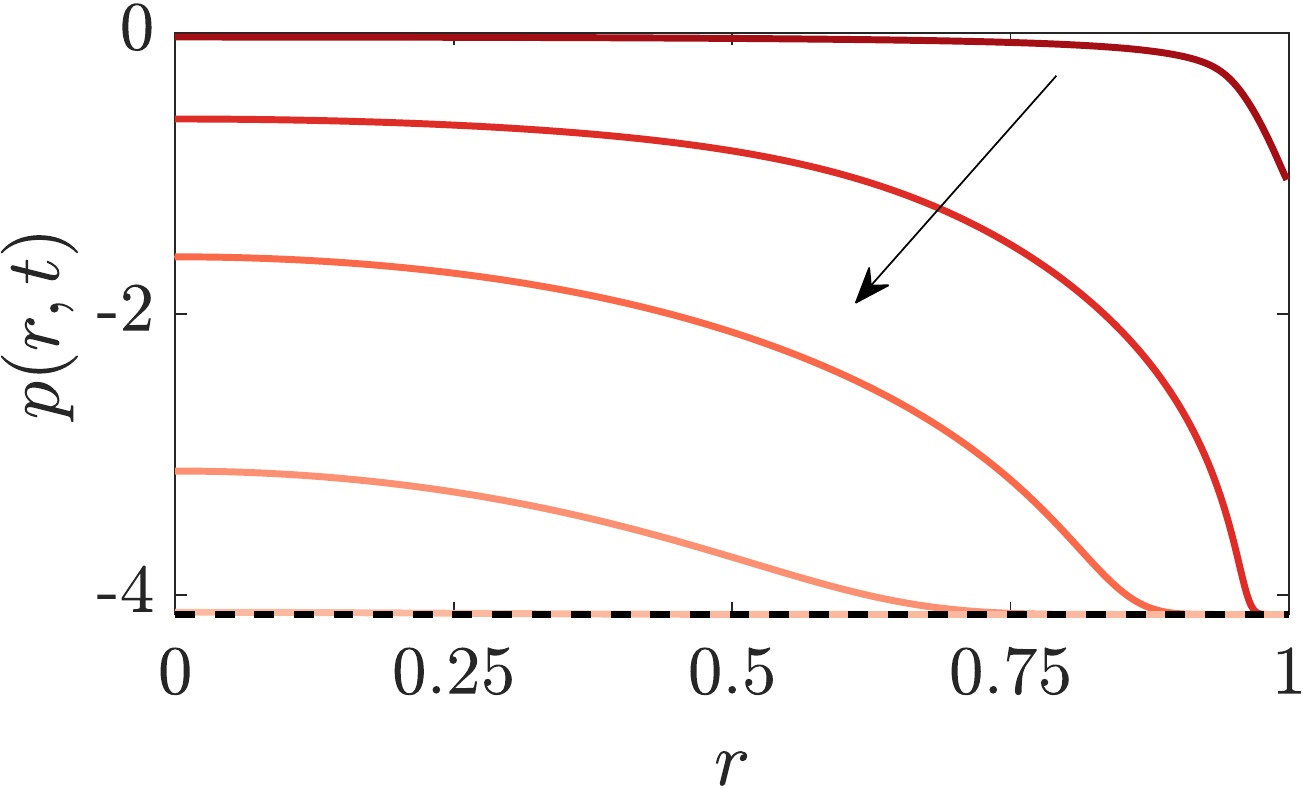}}
  \subfigure[]{\includegraphics[width=0.32\textwidth]{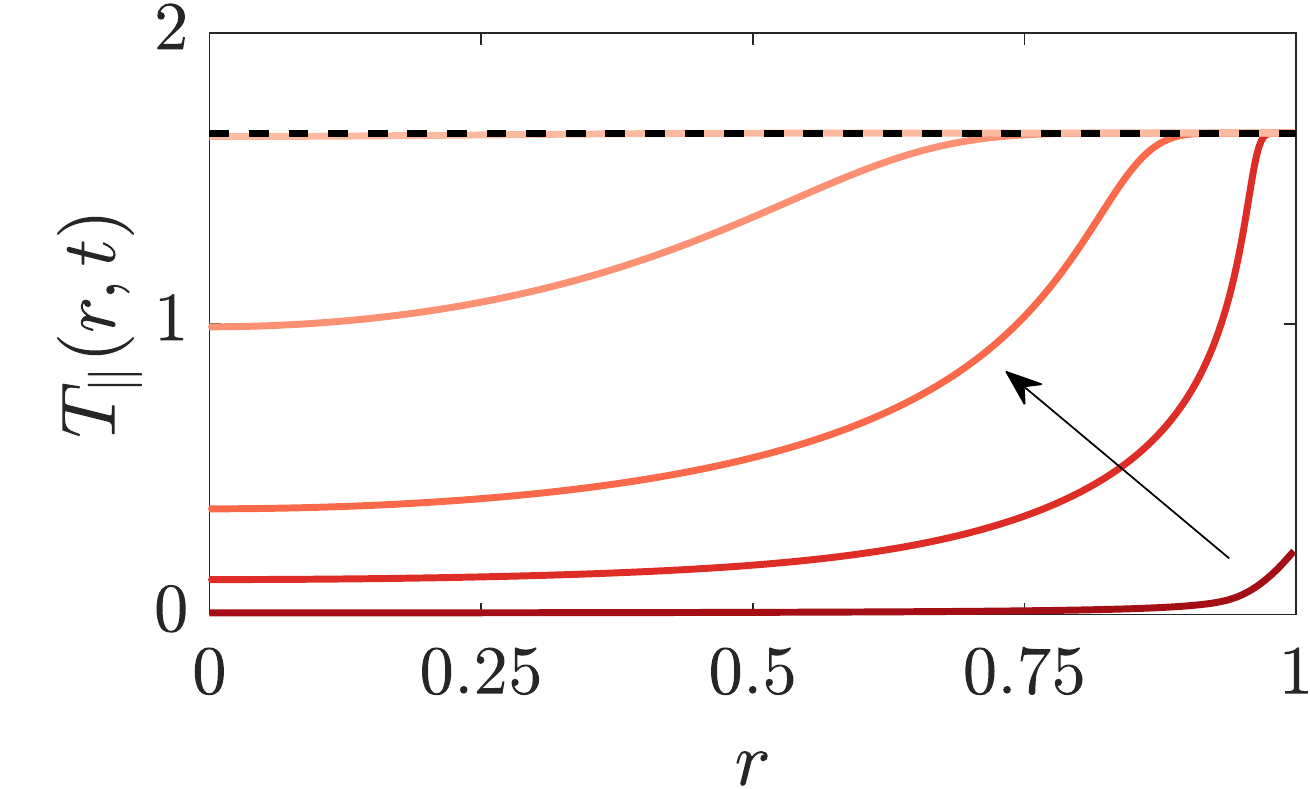}}
  \subfigure[]{\includegraphics[width=0.32\textwidth]{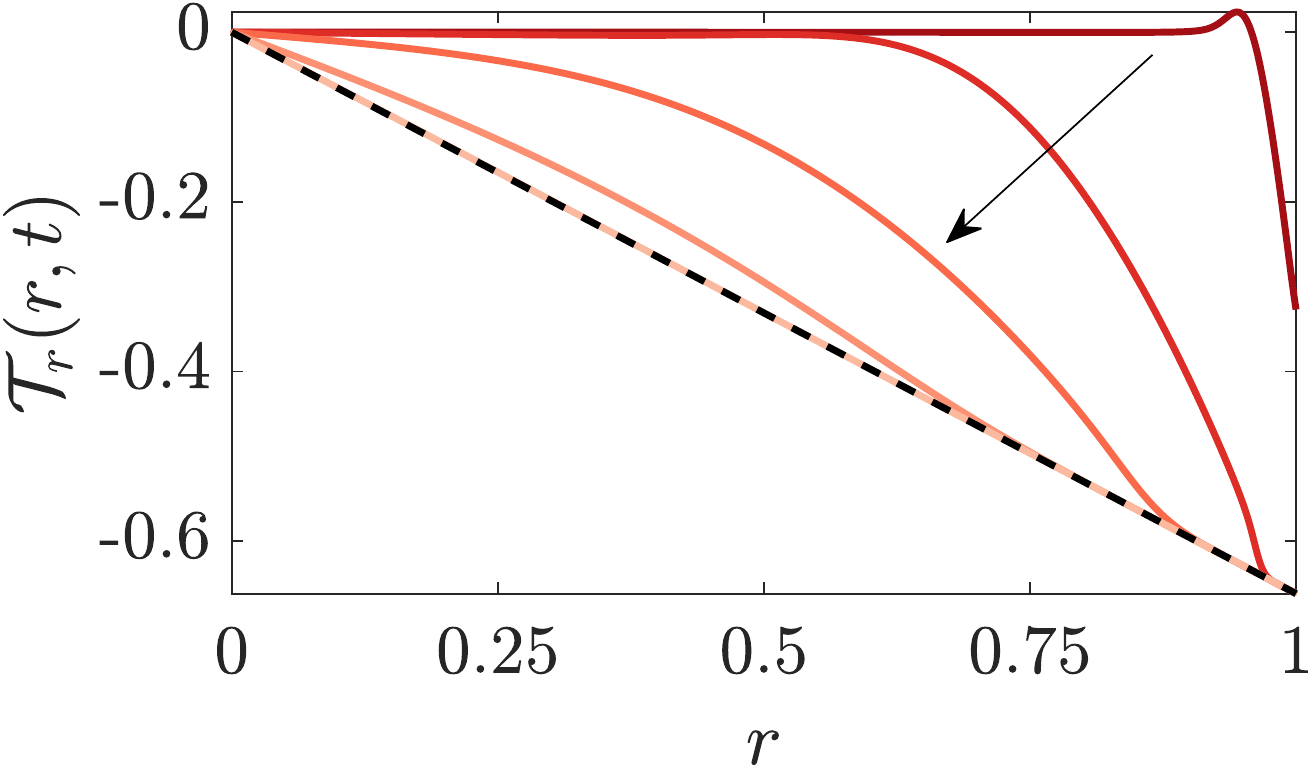}}
  \subfigure[]{\includegraphics[width=0.32\textwidth]{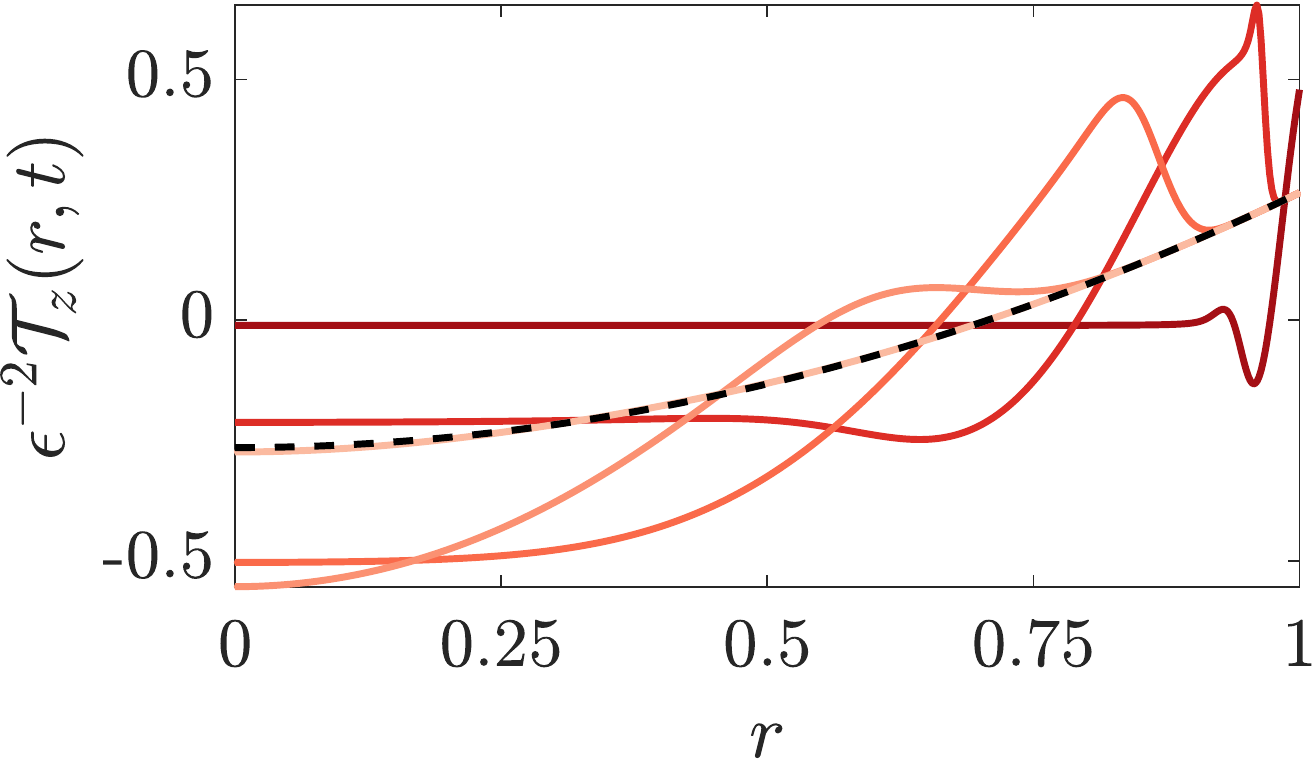}}
  \caption{The dynamics of a drying poroelastic drop at a moderate
    evaporation rate ($\mathcal{Q} = 1$). The spatio-temporal evolution of the (a) fluid fraction (porosity), 
    (b) drop thickness,  (c) pressure,
    (d) total in-plane stress, (e) radial and (f) vertical traction on the
    substrate. The parameter values are $\nu = 0.45$, $\phif_0 = 0.80$,
    and $q(\phif) \equiv 1$.  The solutions are shown at times
    $t = 0.01$, $0.2$, $0.5$, $1$, and $2$. Arrows show the direction of time;
    the dashed black lines denote the steady states.}
  \label{fig:run1}
\end{figure*}

The poromechanics that occur for larger P\'eclet numbers are explored
via numerical simulations of the thin-film equation \eqref{eqn:thin_film_sys} in an
axisymmetric geometry. The initial profile of the drop is assumed to be
parabolic; thus, we take $h_0(r) = 1 - r^2$. For simplicity, the non-dimensional
evaporative mass flux is taken to be a constant, $q(\phif) \equiv 1$.
As a result, all of the fluid will evaporate from the pores of the
solid. The consequences of this simplifying assumption on the dynamics will
be discussed below.

The first case we consider corresponds to a moderate rate of
evaporation with $\mathcal{Q} = 1$. For P\'eclet numbers $\mathcal{Q}$
that are $O(1)$ in size, the time scale of fluid depletion due to evaporation
is commensurate with the time scale of fluid replenishment due to bulk
transport. Thus, the generation of composition gradients within the material
is to be expected.
The initial fluid fraction, or porosity, is set to $\phif_0 = 0.80$
and the Poisson's ratio is taken to be $\nu = 0.45$.
The evolution of the fluid fraction indicates there is a rapid loss of fluid near the contact line, which results
in a completely collapsed (fluid-free) solid; see Fig.~\ref{fig:run1}~(a). 
Due to the sharp decrease in the permeability with the porosity, fluid
from the bulk is prohibited from replenishing that which is lost due to
evaporation. As the drying process continues, a depletion front invades the
drop from the contact line while the fluid content in the bulk decreases
with a weak composition gradient. 
The motion of the depletion front can be detected in the evolution of the
drop thickness. Upstream of the front, the drop thickness remains
stationary because it has converged to its steady-state profile,
while downstream of the front, the drop thickness continues to decrease
as fluid is removed from the pore space; see Fig.~\ref{fig:run1}~(b).

The formation of a depletion front plays a significant role
in the mechanical response of the drop. The localised removal of fluid
from regions near the contact line triggers a vertical compression of the
solid skeleton and leads to a large decrease in the pressure,
as seen in Fig.~\ref{fig:run1}~(c). This zone
of negative pressure propagates into the bulk following the
depletion front. In turn, the negative pressure generates
tensile stresses $\Txx = T_{\parallel} \Ixx$
in both the radial and orthoradial
directions that increasingly penetrate into the bulk with time; 
see Fig.~\ref{fig:run1}~(d). 

The radial traction
can be expressed as
$\mathcal{T}_r = \pdf{(h T_\parallel)}{r}$, which is simply the gradient
of the vertically integrated, total radial stress.
Thus, the behaviour of the radial traction largely mirrors
that of the radial stress: a sharp
gradient develops near the contact line and propagates inwards, as
shown in Fig.~\ref{fig:run1}~(e).  However,
unlike the radial stress, the radial traction settles into a non-uniform
steady state due to the gradients in the drop thickness. 
The negative values of the radial
traction imply that the substrate is being pulled towards the drop
centre. 
The vertical
traction $\TTz$ exhibits non-monotonic behaviour in time, which is likely due to the same
competition between the decrease in drop height and generation of
elastic stress that is captured in Eqn~\eqref{slow:Tz}. 
Regions 
near the contact line experience an upwards force ($\TTz > 0$) that acts to
pull the drop off the substrate and trigger delamination, whereas central
regions of the drop push on the substrate ($\TTz < 0$) and enhance its adhesion; see Fig.~\ref{fig:run1}~(f).

\begin{figure*}
\centering
\subfigure[]{\includegraphics[width=0.32\textwidth]{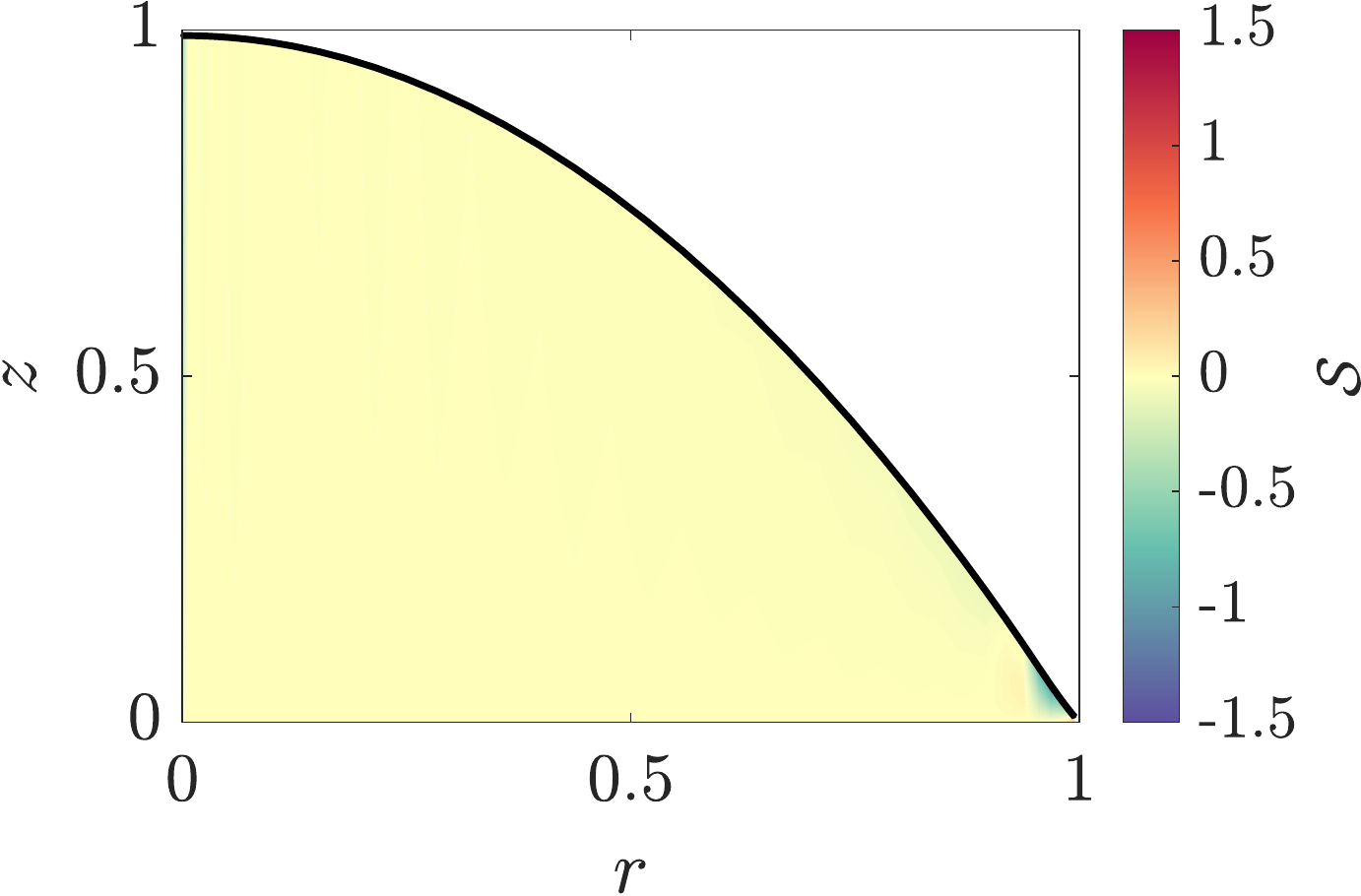}}
\subfigure[]{\includegraphics[width=0.32\textwidth]{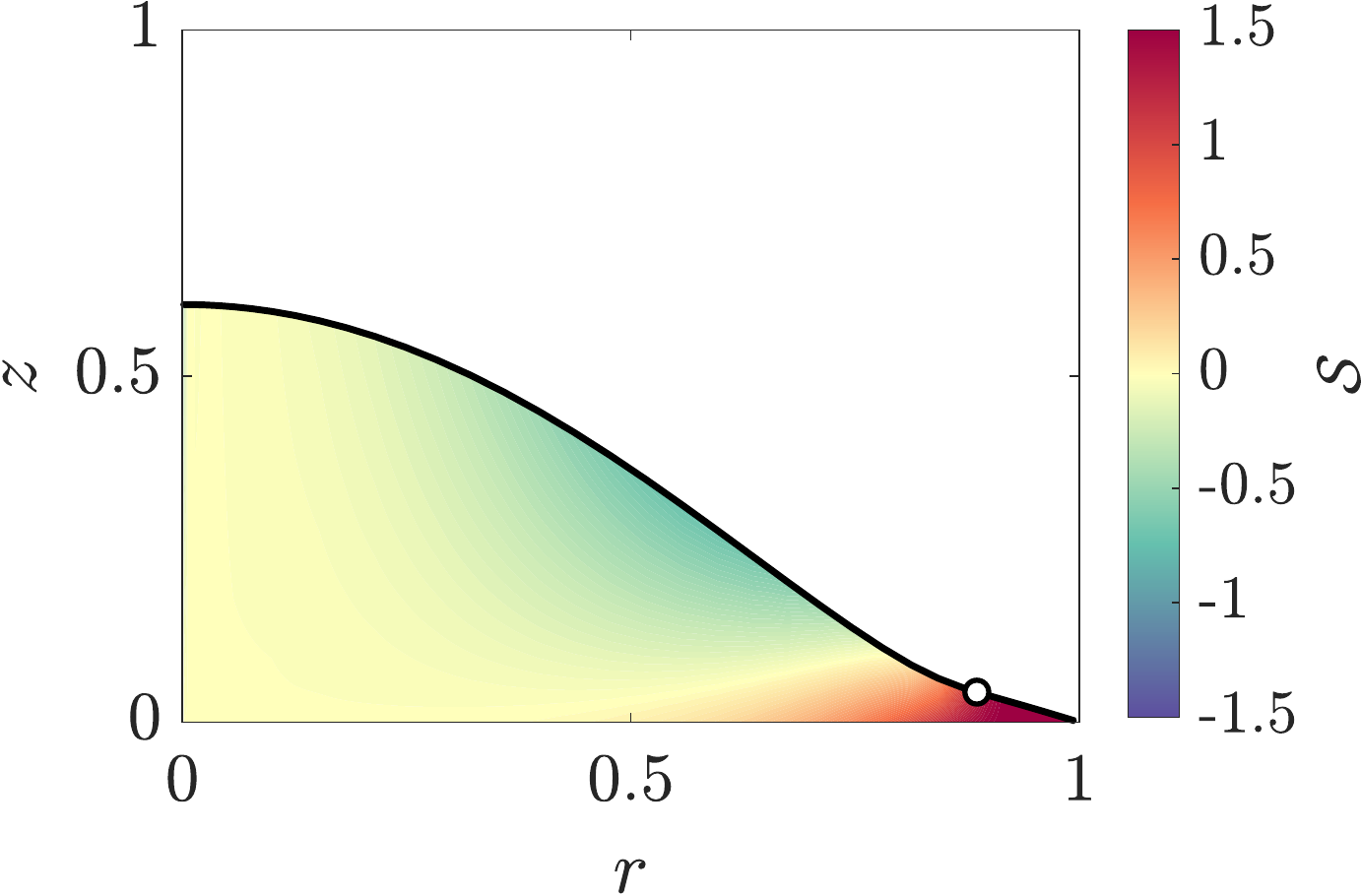}}
\subfigure[]{\includegraphics[width=0.32\textwidth]{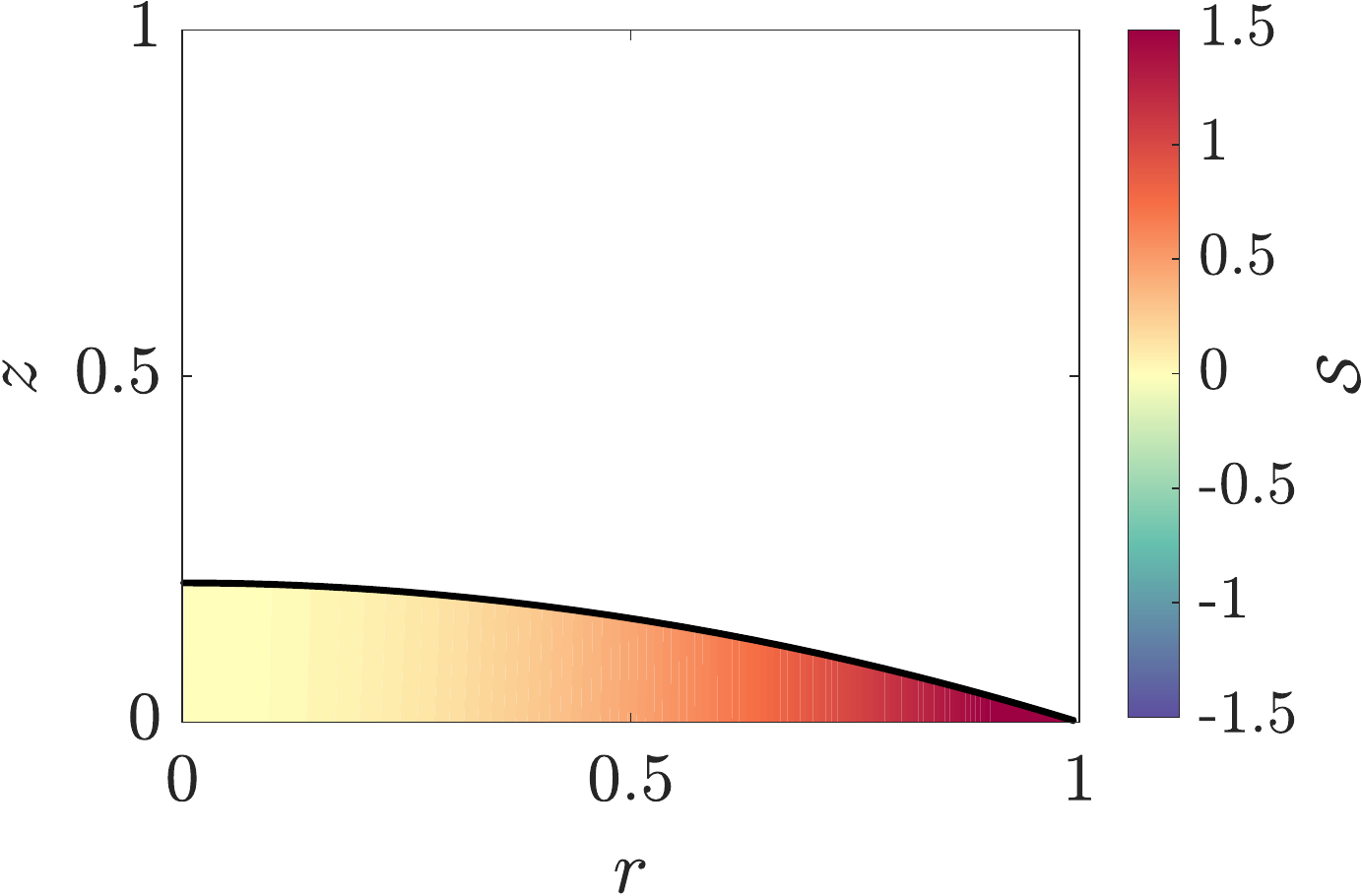}}
\caption{Moderate evaporation rates
  can lead to the hoop stress exceeding
the radial stress in initially parabolic drops.  The rescaled
difference between the radial and hoop stress $\mathcal{S} = \epsilon^{-2}(1+\nu)(\sigma_{rr} - \sigma_{\theta \theta})$ defined by \eqref{eqn:S}
is shown as a heatmap at times (a) $t = 0.01$, (b) $t = 0.5$, and (c) $t = 2$.  The drop profile $h(r,t)$ is shown as the solid black line.  The white circle in (b) depicts the position of the depletion front $r_f(t)$, defined by $\phif(r_f(t),t) = 0.01$. The
parameter values are $\mathcal{Q} = 1$,
$\nu = 0.45$, $\phif_0 = 0.80$, and $q(\phif) = 1$.}
\label{fig:run1_S}
\end{figure*}

Larger evaporation rates drive the emergence of non-parabolic drop shapes that
result in the hoop stress exceeding the radial stress.
By computing the rescaled stress
difference $\mathcal{S}$ given by \eqref{eqn:S}, we find that a localised
region
appears at the contact line where the hoop stress exceeds the
radial stress ($\mathcal{S} < 0$); see Fig.~\ref{fig:run1_S}~(a).
This localised is referred to as the ``hoop zone''.
As time increases, the hoop zone propagates along the free surface of the drop towards the centre, and, at the contact line, the radial stress overtakes the
hoop stress ($\mathcal{S} > 0$); see Fig.~\ref{fig:run1_S}~(b).  As the drop completely dries out,
the hoop zone dissipates and the radial stress dominates across the entirety of the drop, as seen in Fig.~\ref{fig:run1_S}~(c).  The propagating depletion front
separates the hoop zone from the region dominated by the radial stress.  
The region behind (upstream of) the depletion front is fluid-free and the
drop profile is simply a rescaled version of the initial parabolic profile, $h(r,t) \simeq (1 - \phif_0) h_0(r)$; thus, the radial stress dominates, in accordance with
\eqref{eqn:S0}. Ahead (downstream) of the depletion front, the drop profile
becomes non-parabolic due to the non-uniform removal of fluid, thus leading
to the hoop zone.

\begin{figure*}
  \centering
  \subfigure[]{\includegraphics[width=0.32\textwidth]{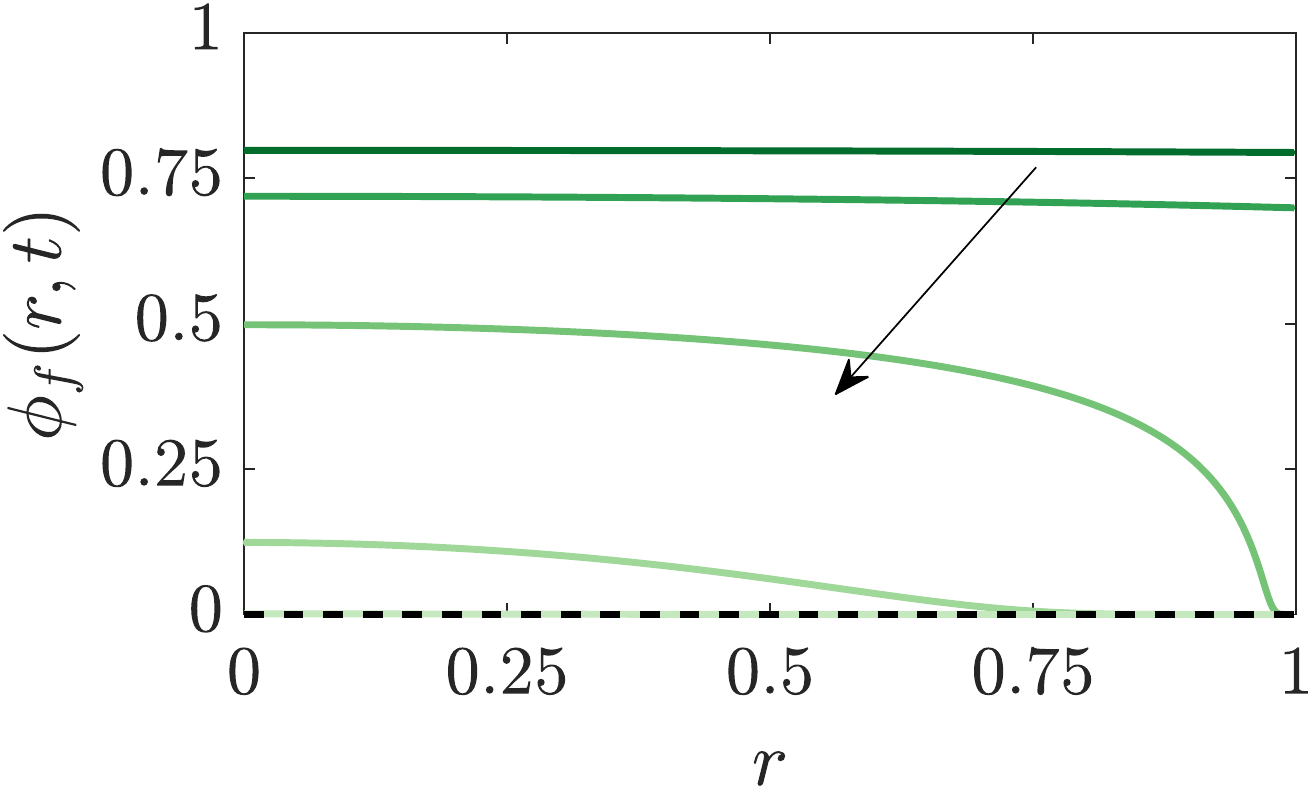}}
  \subfigure[]{\includegraphics[width=0.32\textwidth]{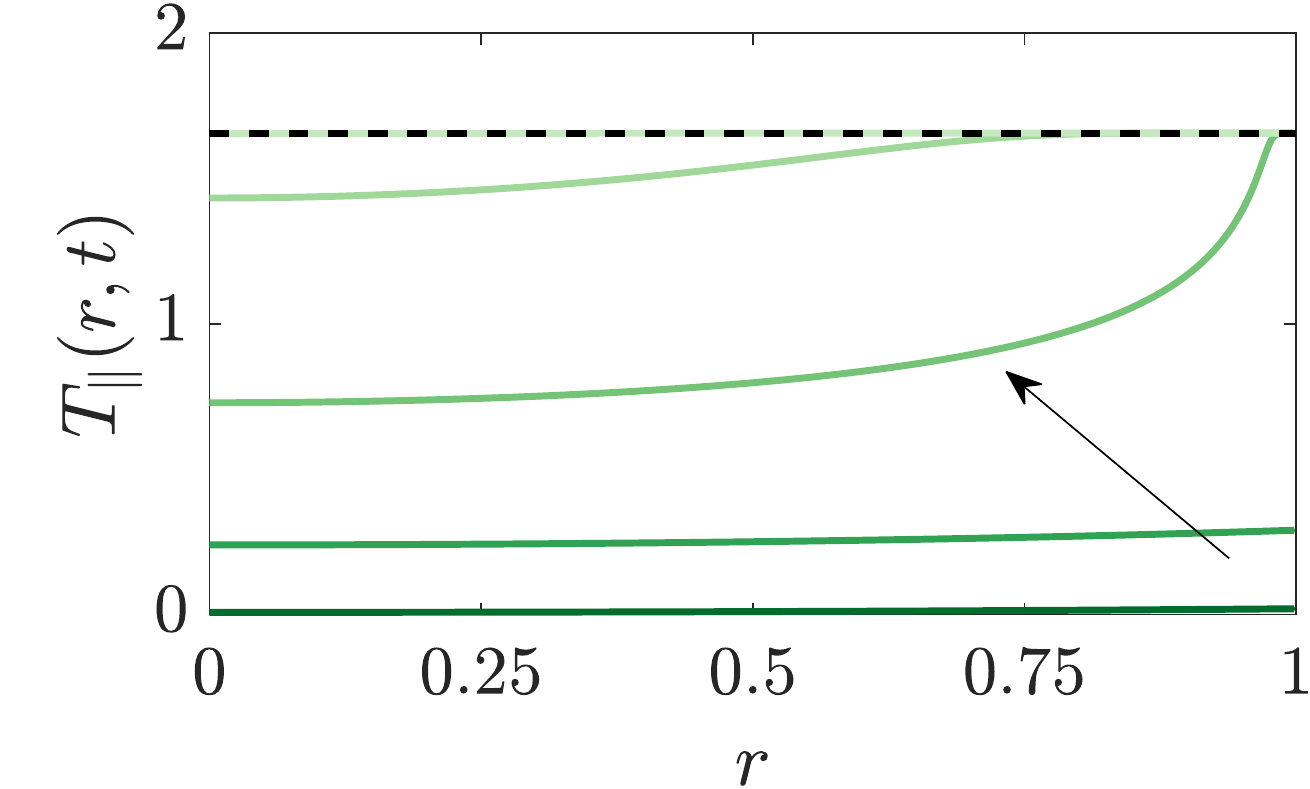}}
  \subfigure[]{\includegraphics[width=0.32\textwidth]{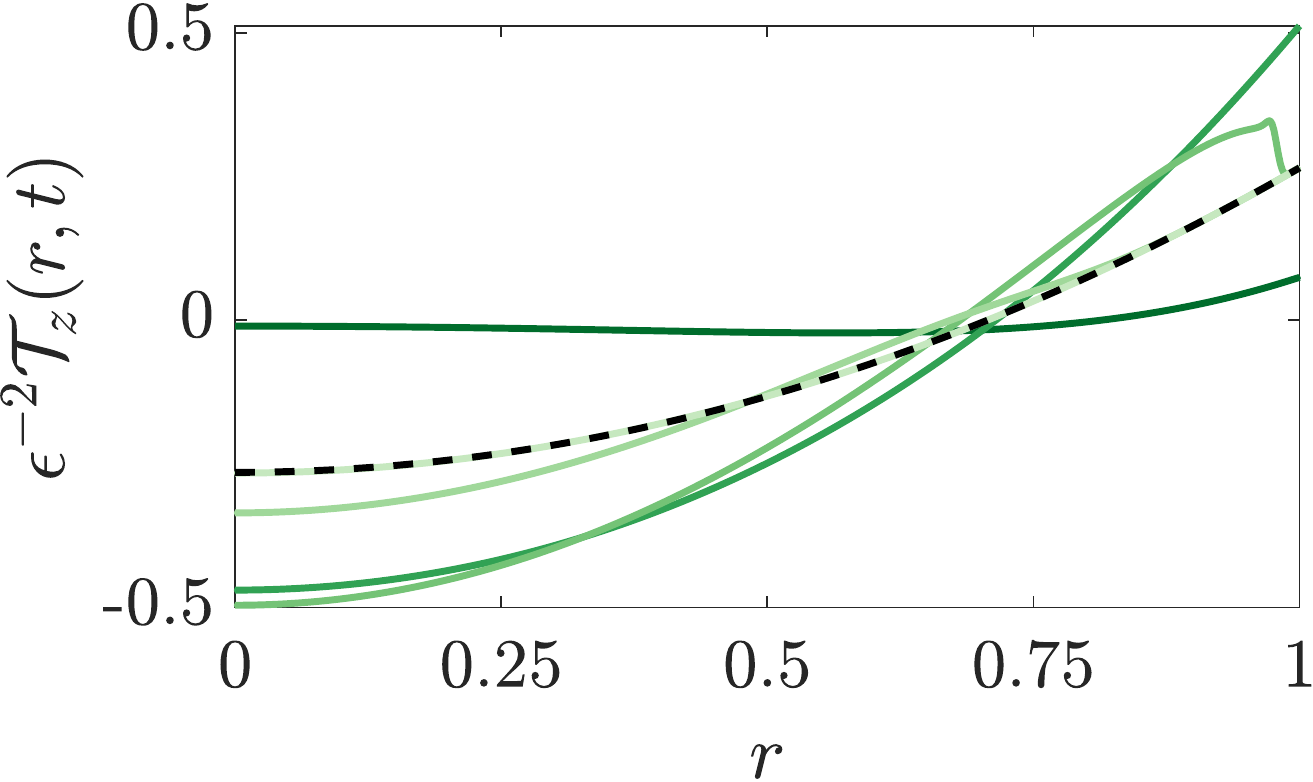}}
  \subfigure[]{\includegraphics[width=0.32\textwidth]{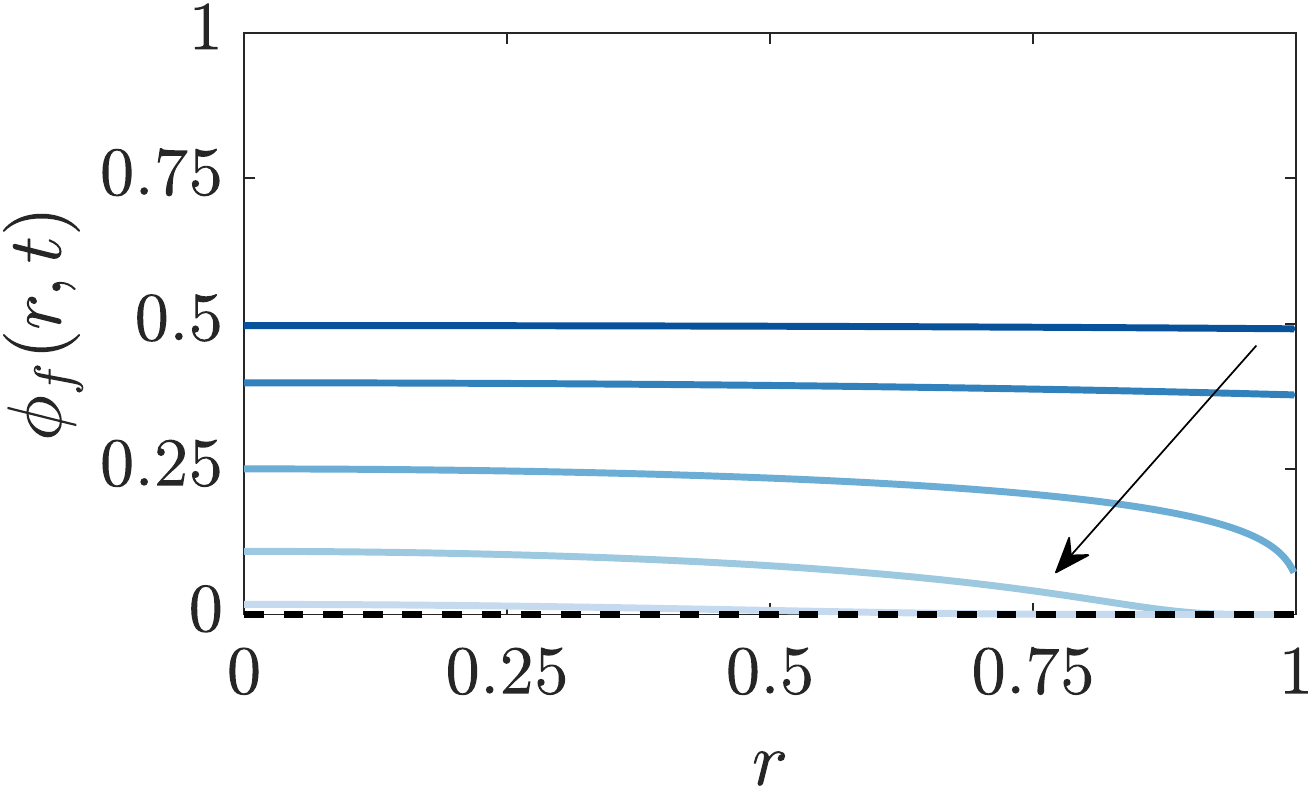}}
  \subfigure[]{\includegraphics[width=0.32\textwidth]{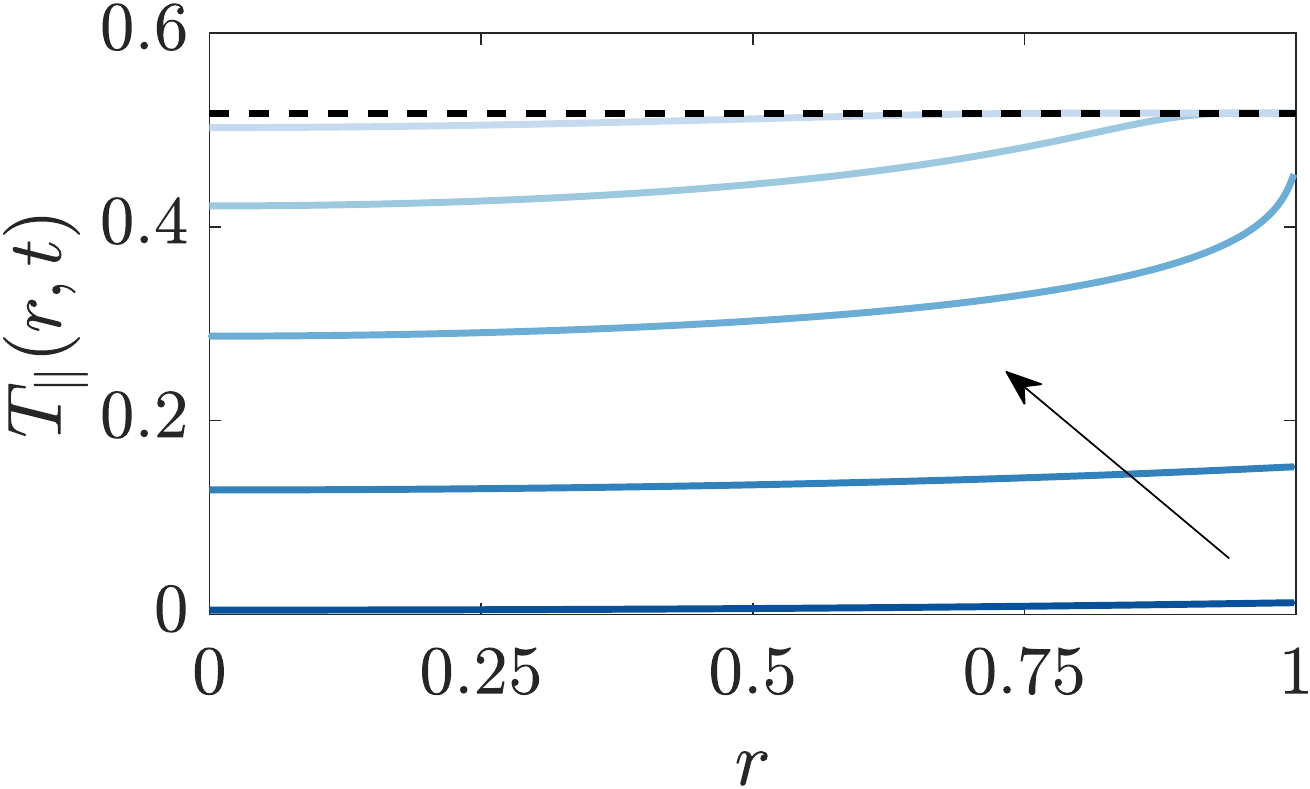}}
  \subfigure[]{\includegraphics[width=0.32\textwidth]{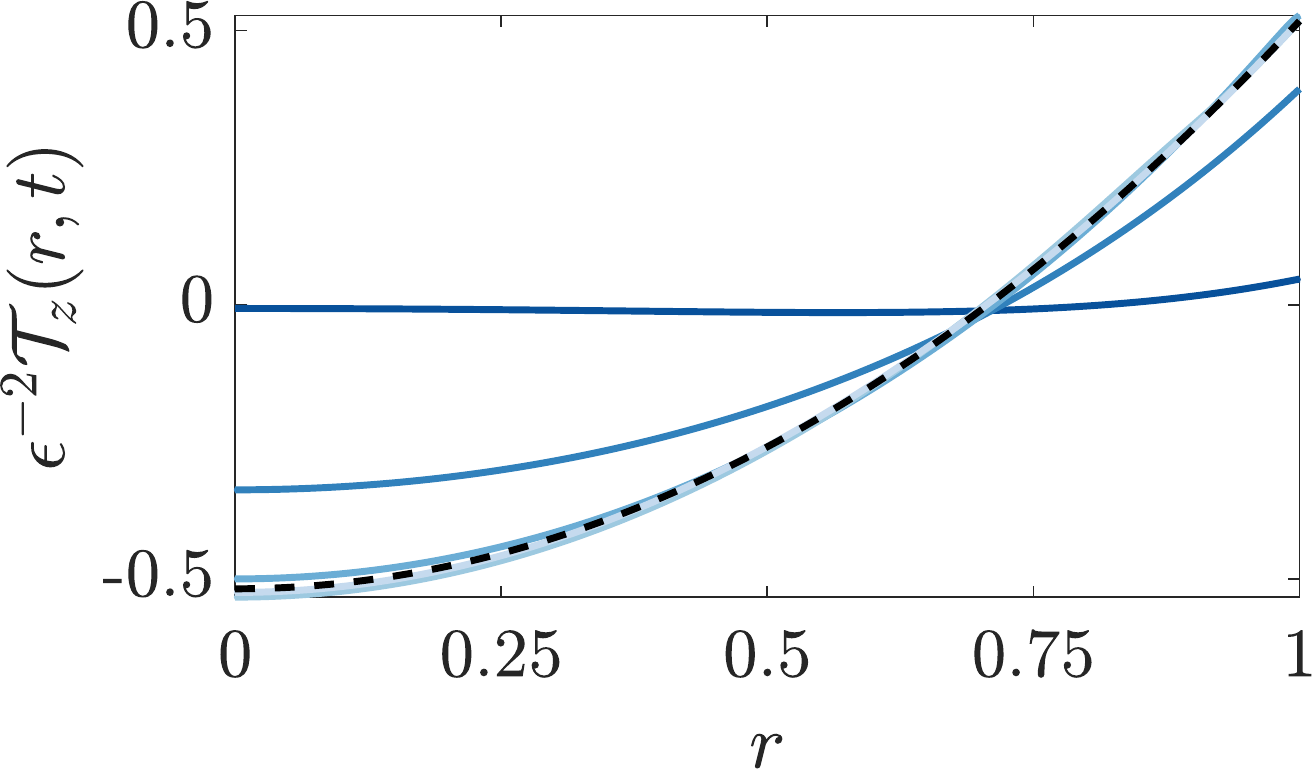}}
  \caption{Drop dynamics with a small evaporation rate ($\mathcal{Q} = 0.1$). In panels (a)--(c), the initial
    fluid fraction is $\phif_0 = 0.80$. In panels (d)--(f), the initial
    fluid fraction is $\phif_0 = 0.50$. In all panels, $\nu = 0.45$,
    $q(\phif) \equiv 1$, 
    and the solutions are shown at times
    $t = 0.1$, $2$, $5$, $10$, and $20$. Arrows show the direction of time;
    the dashed black lines denote the steady states.}
  \label{fig:run23}
\end{figure*}

We now turn our attention to the drop dynamics that occur
for slower rates of evaporation
by considering the case when $\mathcal{Q} = 0.1$. We first consider a drop
with the same parameters as in Fig.~\ref{fig:run1} by setting
$\nu = 0.45$ and $\phif_0 = 0.80$. 
The fluid fraction initially
decreases while remaining approximately uniform throughout the drop,
see Fig.~\ref{fig:run23}~(a), which is consistent with the findings of the slow-evaporation limit.  The homogeneous drying of the drop gives rise
to roughly uniform in-plane stresses as well, as shown in Fig.~\ref{fig:run23}~(b).  The in-plane stresses, in turn, generate
a vertical traction with a roughly parabolic initial profile; see Fig.~\ref{fig:run23}~(c). 
Eventually, the loss of fluid triggers a sharp
decrease in the permeability. Weak gradients in the fluid fraction near the
contact line are amplified, resulting in a propagating depletion
front; see Fig.~\ref{fig:run23}~(a). 

Decreasing the
initial fluid fraction to $\phif_0 = 0.5$ and keeping the other parameters fixed leads to qualitatively similar dynamics to those seen in
Fig.~\ref{fig:run23}~(a)--(c).  However, in this case,
the depletion front is more diffuse (Fig.~\ref{fig:run23}~(d))
and the in-plane stresses $T_\parallel$ are smaller (Fig.~\ref{fig:run23}~(e))
due to the solid skeleton undergoing less volumetric contraction; recall
from \eqref{eqn:J} that $J = 1 - \phif_0$ at the steady state.
The vertical traction
monotonically approaches its steady-state profile (Fig.~\ref{fig:run23}~(g)),
which
is larger in magnitude than the case when $\phif_0 = 0.8$ due to 
the smaller change in drop thickness.

In all of the cases considered so far, it has been assumed that evaporation
completely dries the solid, removing all fluid from within the pore space.
Relaxing this assumption, so that some fluid remains, will curtail the
decrease in the permeability. This will result in weaker gradients
in the fluid fraction and, consequently, in all of the other quantities
as well. Moreover,
it may prohibit the formation of a depletion front altogether.

\subsection{Drops with large contact angle}

The finite element method is used to compute the steady-state
stress distribution
in axisymmetric drops with large contact angles.
Under the steady-state assumption, the full time-dependent
model described in Sec.~\ref{sec:model} reduces to the equilibrium
equations of nonlinear elasticity, $\nabla \cdot \tens{\sigma} = \nabla p$,
with an incompressibility
constraint $J = J_0$. The constant $J_0 < 1$ describes the
volumetric contraction due to drying.
The governing equations are solved in the reference
configuration. The finite element method is implemented with
FEniCS~\cite{logg2012, alnaes2015} using P2-P1
elements for displacement and pressure, respectively.
In all of the simulations, the initial profile
of the drop is taken to 
be a parabola that is represented in dimensional form as
$h_0(r)/R = \epsilon\left(1 - (r/R)^2\right)$.
The initial contact angle satisfies $\tan \varphi_0 = 2 \epsilon$ and thus
$\varphi_0 \sim 2 \epsilon$ for $\epsilon \ll 1$.  In addition, we set
$J_0 = 1/2$, corresponding to a drop that has shed half of its volume.

\begin{figure}
  \centering
  \includegraphics[width=\columnwidth]{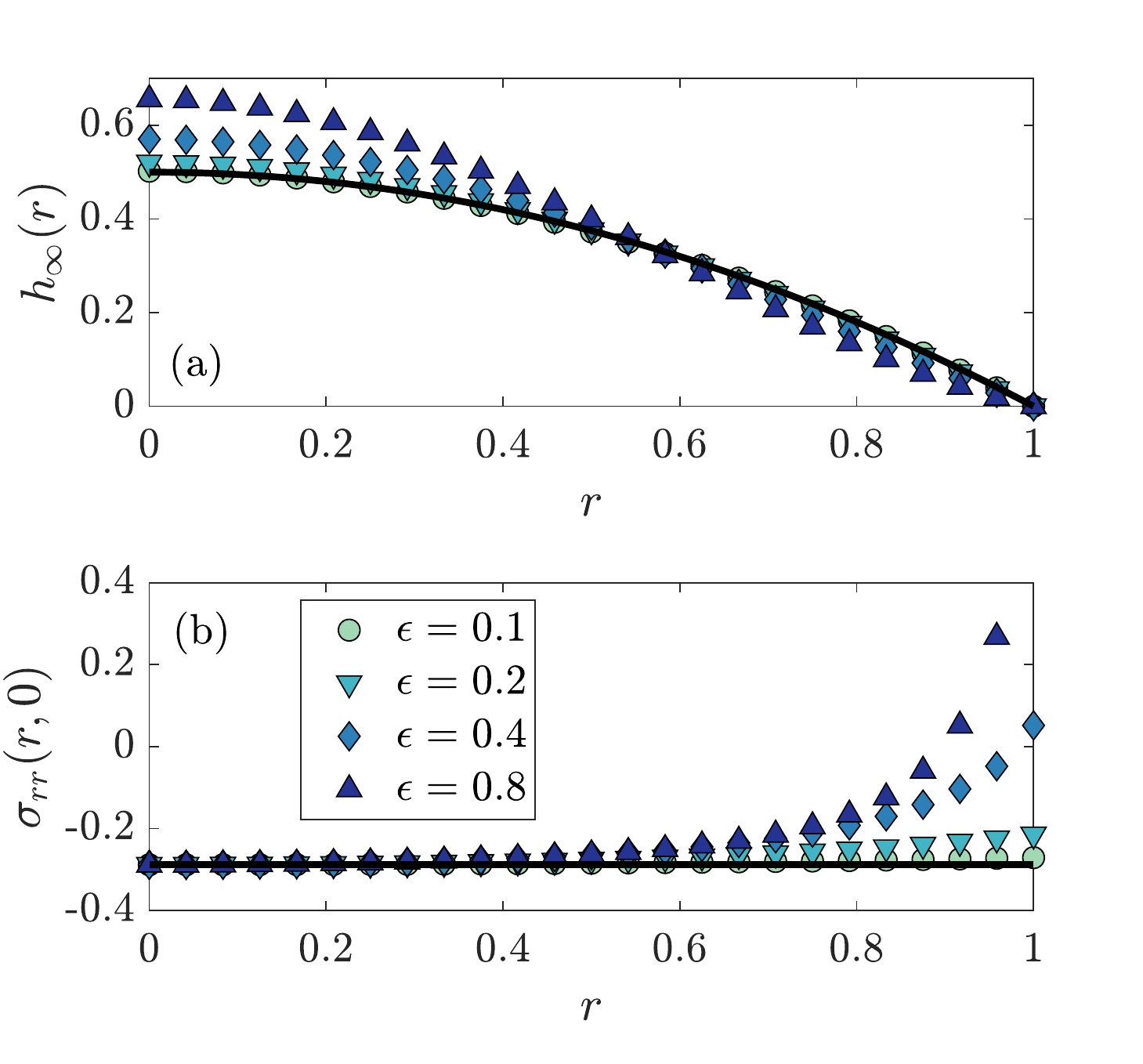}
  \caption{The (a) equilibrium drop thickness $h_\infty$ and (b)
    radial elastic stress along the substrate for different drop aspect ratios
    $\epsilon = H / R$. Symbols denote quantities computed using the finite
    element method. The solid black lines denote asymptotic solutions. The
  legend applies to both panels.}
  \label{fig:fem_compare}
\end{figure}

We first compute the steady-state drop thickness $h_\infty$ for a
range of aspect ratios $\epsilon =  H / R$. When
$\epsilon = 0.1$ and $0.2$ ($\varphi_0 = 11^\circ$ and $22^\circ$),
the drops have parabolic profiles that are in good agreement the asymptotic
theory; see Fig.~\ref{fig:fem_compare}~(a). However, as $\epsilon$ increases
to 0.4 and then to 0.8 ($\varphi_0 = 39^\circ$ and $58^\circ$),
deviations from a parabolic profile begin to emerge.
The profile that arises when $\epsilon = 0.8$
closely resembles that seen by Pauchard and Allain~\cite{pauchard2003}
when studying drying colloidal drops with contact angles
on the order of 45$^\circ$.  When the contact angle is large,
the solid skeleton is generally further away from the substrate and hence
less influenced by the no-slip (perfect adhesion) condition. Thus,
drying leads to greater radial displacements. However, the radial
displacement is constrained near $r = 0$ due to the assumption
of axisymmetry. The net result is that solid near the contact
line is displaced inwards and, in order to conserve solid volume, the vertical
contraction of the drop near the centre is reduced.


Increasing the contact angle also leads to marked changes in the radial
elastic stress $\sigma_{rr}$. When $\epsilon = 0.1$ and $0.2$, the
radial elastic stress along the substrate is compressive and nearly uniform,
in agreement with the asymptotic solutions; see Fig.~\ref{fig:fem_compare}~(b).
However, increasing $\epsilon$ leads to larger gradients and the emergence
of a region near the contact line where the radial elastic stress becomes
tensile. Explicitly calculating $\sigma_{rr}$ along the substrate
reveals that its tensile nature is a nonlinear effect arising from large
shear strains $\pdf{u_r}{z}$.

\begin{figure*}
  \centering
  \subfigure[]{\includegraphics[width=0.32\textwidth]{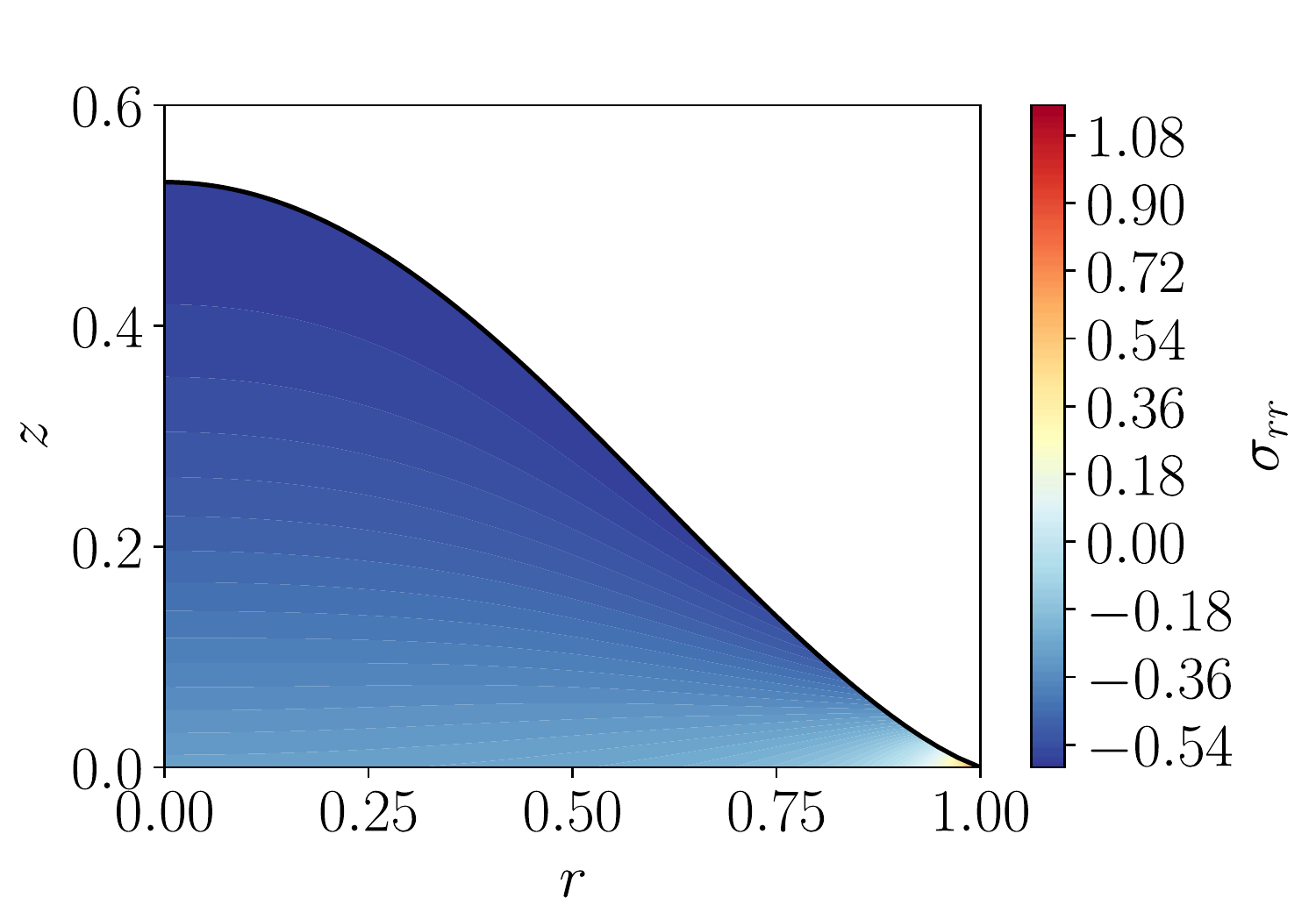}}
  \subfigure[]{\includegraphics[width=0.32\textwidth]{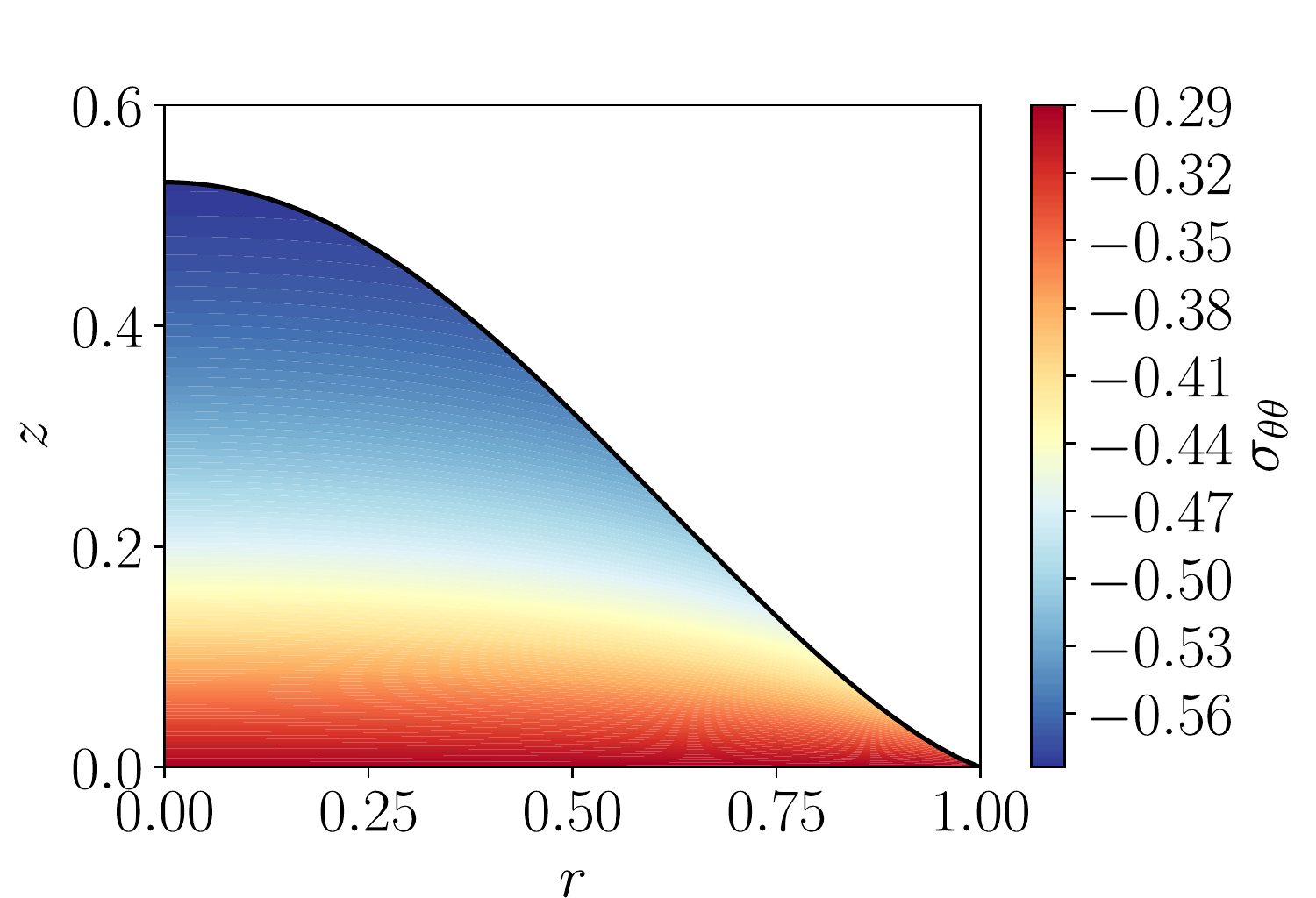}}
  \subfigure[]{\includegraphics[width=0.32\textwidth]{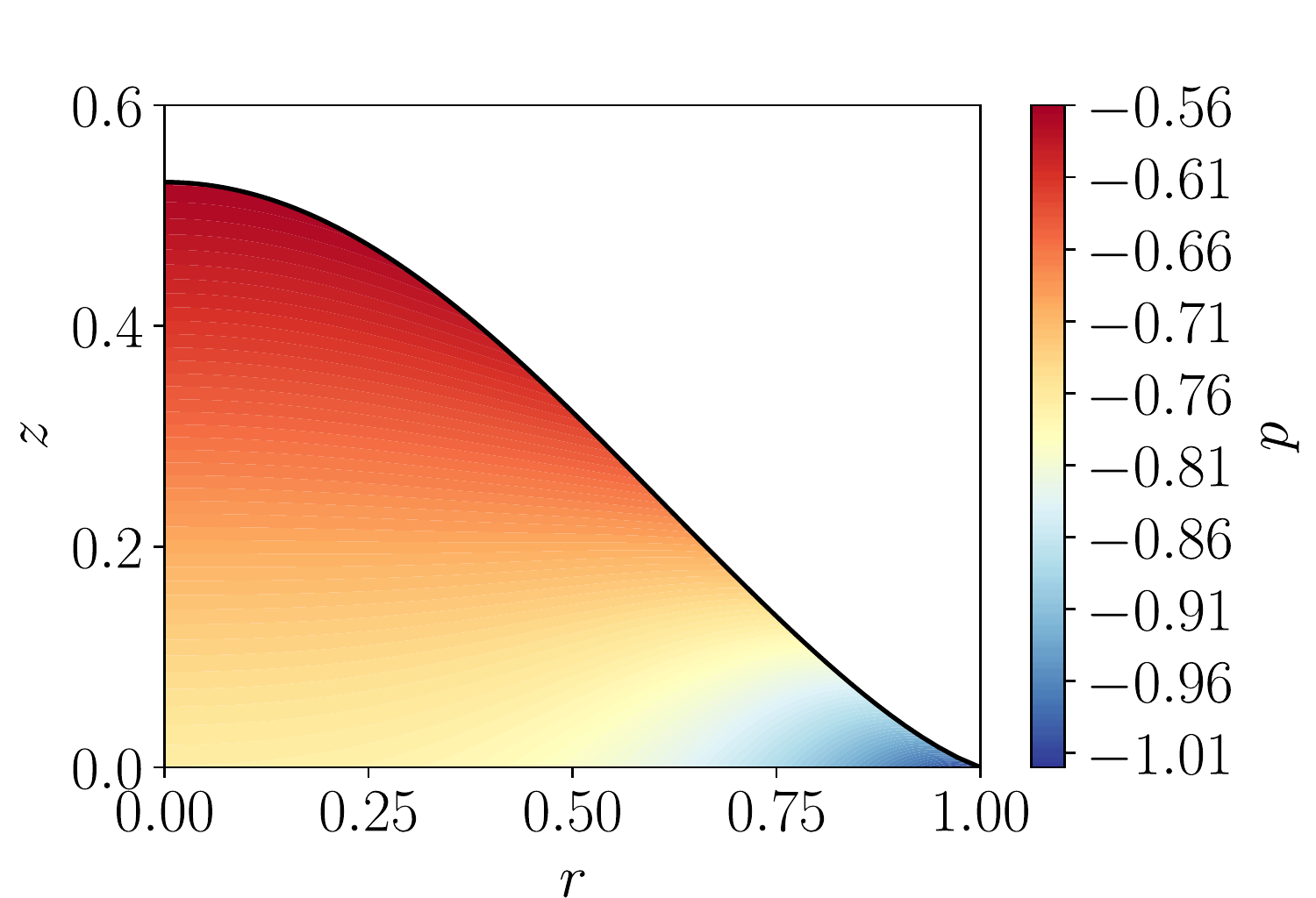}}
  \caption{The (a) radial elastic stress, (b) orthoradial
    elastic stress, and (c) pressure
    in a dried poroelastic drop with large initial contact angle. The drop
    has shed half of its volume due to fluid loss ($J = 0.50$).
    We set $\epsilon = 0.80$ ($\varphi_0 \simeq 58^\circ$) and $\nu = 0.30$.}
  \label{fig:fem_stress}
\end{figure*}

To further explore the poromechanics of drying with large contact angles,
we have computed the spatial distribution of the radial and
orthoradial elastic stresses, along with the
pressure, in a drop with an aspect ratio of $\epsilon = 0.8$.
The radial elastic stress is generally compressive, with the exception of
a small tensile region near the contact line; see
Fig.~\ref{fig:fem_stress}~(a). The orthoradial elastic stress is also
compressive; see Fig.~\ref{fig:fem_stress}~(b). The magnitude of the
orthoradial elastic stress increases with distance from the substrate due to
the radial displacement increasing in
magnitude as well. The pressure is negative throughout
the drop and is concentrated near the contact line;
see Fig.~\ref{fig:fem_stress}~(c). Computing the total (Cauchy)
stress by subtracting the pressure from the elastic stresses
shows that the drop is under tension in both the radial and orthoradial
directions. However, the radial stress exceeds the orthoradial stress,
particularly at locations near the contact line.

The stress profiles shown in Fig.~\ref{fig:fem_stress} indicate that
many of the conclusions obtained from the
asymptotically reduced model still apply when the contact angle of the
drop is not small. However, from Fig.~\ref{fig:fem_compare}, we see that
the quantitative accuracy of the asymptotic reduction can only be ensured
for drops with initial contact angles that are smaller than  $20^\circ$.

As a final point, experiments have shown that the drying pathway for colloidal
drops with large contact angles involves the formation of an elastic
skin at the free surface~\cite{zang2019}.
Drying-induced stresses lead to buckling of the skin~\cite{pauchard2003}
as opposed to fracture. Extending the poroelastic model
proposed here to shell-like geometries would allow drying-induced buckling
patterns to be studied.




\section{Concluding remarks}
\label{sec:conclusion}

By combining nonlinear poroelasticity with the lubrication approximation, we have derived a simplified model that offers new insights into the generation of mechanical stress during the drying of complex drops.
The asymptotic analysis indicates that the initial profile of the solid skeleton $h_0$ plays a central role in the poromechanics of drop drying,
as it controls the in-plane motion of the solid skeleton. Using the asymptotic
solutions for the stress, it is possible to predict the alignment of
desiccation fractures.

A limitation of the model proposed here is that is based on the
assumption that the drop has a pre-existing poroelastic structure.
That is, the model does not
consider the regime in which the drop is liquid. As a consequence, the
initial profile of the solid skeleton $h_0$
must be provided as input to the model.
An important
area of future work is to develop an extended model that captures the
fluid mechanics of drying and the sol-gel transition, 
with the aim of predicting $h_0$.
Routh and Russel~\cite{routh1998} studied a similar problem but assumed the
porous solid was rigid.

When modelling the drying of biological fluids before gelation occurs,
non-Newtonian effects may be important to consider.
For example, blood is often
modelled as a Carreau--Yasuda fluid~\cite{boyd2007}.
In this case, the relevance of
non-Newtonian effects can be assessed through the quantity $(\lambda \dot{\gamma})^a$, where $\lambda$ is a relaxation time, $\dot{\gamma}$ is the shear rate,
and $a$ is a constant. Abraham \etal\cite{abraham2005}
report that $\lambda = 8.2$~s and
$a = 0.64$ for blood. For a thin drop in the lubrication limit,
$\dot{\gamma} \sim  U / (\epsilon R)$.
Sobac and Brutin~\cite{sobac2014} state that, before gelation,
the fluid velocity $U$ is dominated by capillary action and provide a value of
$U = 8$~$\mu$m s\unit{-1}. Using the parameter values in
Sec.~\ref{sec:params} gives $(\lambda \dot{\gamma})^a \sim 0.26$,
which is small but not negligible.
After gelation, the liquid component of blood (mainly water) will flow through
a porous network composed of solid biological components
(mainly red blood cells).
Thus, describing the macroscopic flow field using Darcy's law, as done
here, is appropriate.

With a satisfactory initial profile for the solid skeleton, the asymptotic approach developed here can be extended to a wide range of new problems that capture, for example, delamination, substrate deformability, and fracture.
These problems will help to unravel the complex interplay between
physical mechanisms that govern the drying of complex fluids and provide
a deeper understanding of the various modes of mechanical instability
that can occur.


\begin{acknowledgements}

We thank Aran Uppal, Ludovic Pauchard, Irmgard Bischofberger, and Paul Lilin for stimulating discussions about pattern formation in drying colloidal drops. 

\end{acknowledgements}


\bibliography{refs}

\end{document}